\documentclass[review]{elsarticle}

\usepackage{lineno,hyperref}
\modulolinenumbers[5]

\journal{Journal of Computational Physics}

\usepackage{graphicx}
\usepackage[english]{babel}
\usepackage{amssymb}
\usepackage{amsbsy}
\usepackage{amsmath}
\usepackage{color}
\usepackage{psfrag,epsfig}

\usepackage{algorithm}
\usepackage{algpseudocode}

\usepackage{times}
\usepackage{mathptm}

\newcommand{\bg}{\boldsymbol{\gamma}}

\def\bg{{\bf g}}
\def\bn{{\bf n}}
\def\bq{{\bf q}}

\def\bK{{\bf K}}
\def\bQ{{\bf Q}}

\def\Frac{\displaystyle \frac}

\newif\ifnotesw \noteswtrue
\newcommand{\jeffrey}[1]{\ifnotesw  \textcolor[rgb]{0,0,1}{  $\spadesuit$Jeffrey:\ {\sf \bf \it #1}\ $\spadesuit$ }\fi}

\noteswfalse   

\usepackage{changes}

\usepackage{siunitx}
\usepackage{bm}
\usepackage{amsfonts}
\usepackage{epstopdf}
\ifpdf
  \DeclareGraphicsExtensions{.eps,.pdf,.png,.jpg}
\else
  \DeclareGraphicsExtensions{.eps}
\fi

\begin{document}

\begin{frontmatter}

\title{Flow and Transport in Three-Dimensional Discrete Fracture Matrix Models using Mimetic Finite Difference on a Conforming Multi-Dimensional Mesh}

\author{Jeffrey D. Hyman \fnref{jeffrey}  }
\fntext[jeffrey]{Corresponding Author: jhyman@lanl.gov}

\author{Matthew R. Sweeney\fnref{matt}  }
\fntext[matt]{msweeney2796@lanl.gov}

\author{Carl W. Gable\fnref{carl}  }
\address{Computational Earth Science (EES-16), Earth and Environmental Sciences Division, Los Alamos National Laboratory, Los Alamos New Mexico, USA}
\fntext[carl]{gable@lanl.gov}

\author{Daniil Svyatsky \fnref{daniil}  }
\fntext[daniil]{dasvyat@lanl.gov}

\author{Konstantin Lipnikov \fnref{konstantin}  }
\fntext[konstantin]{lipnikov@lanl.gov}

\author{J. David Moulton \fnref{david}  }
\address{Applied Mathematics and Plasma Physics (T-5), Theoretical Division, Los Alamos National Laboratory, Los Alamos New Mexico, USA }
\fntext[david]{moulton@lanl.gov}

\begin{abstract}
We present a comprehensive workflow to simulate single-phase flow and transport in fractured porous media using the discrete fracture matrix approach.
The workflow has three primary parts: (1) a method for conforming mesh generation of and around a three-dimensional fracture network, (2) the discretization of the governing equations using a second-order mimetic finite difference method, and (3) implementation of numerical methods for high-performance computing environments. 
A method to create a conforming Delaunay tetrahedralization of the volume surrounding the fracture network, where the triangular cells of the fracture mesh are faces in the volume mesh, that addresses pathological cases which commonly arise and degrade mesh quality is also provided.
Our open-source subsurface simulator uses a hierarchy of process kernels (one kernel per physical process) that allows for both strong and weak coupling of the fracture and matrix domains.
We provide verification tests based on analytic solutions for flow and transport, as well as numerical convergence.
We also provide multiple expositions of the method in complex fracture networks. 
In the first example, we demonstrate that the method is robust by considering two scenarios where the fracture network acts as a barrier to flow, as the primary pathway, or offers the same resistance as the surrounding matrix. 
In the second test, flow and transport through a three-dimensional stochastically generated network containing 257 fractures is presented. 
\end{abstract}

\begin{keyword}
Discrete Fracture Matrix, Computational Geometry, Mimetic Finite Difference, High Performance Computing, Flow and Transport in Fractured Media, Discrete Fracture Network, DFN, DFM, Delaunay Triangulation
\end{keyword}

\end{frontmatter}


\section{Introduction}\label{sec:introduction}

There are several computational models for simulating the flow of fluids and the associated transport of dissolved chemicals through subsurface fractured systems, which is of great importance for understanding the behavior of natural subsurface fracture systems and a variety of industrial and civil engineering endeavors, including the environmental restoration of contaminated fractured media~\cite{national1996rock,neuman2005trends,vanderkwaak1996dissolution}, aquifer storage and management~\cite{kueper1991behavior}, hydrocarbon extraction~\cite{hyman2016understanding,middleton2015shale}, longterm storage of spent civilian nuclear fuel~\cite{hadgu2017comparative,joyce2014multiscale}, and CO$_2$ sequestration~\cite{hyman2019characterizing,jenkins2015state}.
The computational models currently used today can be loosely grouped into three categories~\cite{berre2018flow}: (1) continuum models - including stochastic continuum~\cite{neuman2005trends,neuman1988use,tsang1996tracer} and  dual-porosity / dual-permeability models~\cite{gerke1993dual,lichtner2014modeling,zimmerman1993numerical}, (2) discrete fracture networks (DFN) models~\cite{cacas1990modeling,long1982porous,maillot2016connectivity,makedonska2016evaluating,nordqvist1992variable}, and (3) discrete fracture matrix (DFM) models~\cite{ahmed2015control,ahmed2015three,antonietti2016mimetic,arraras2019mixed,boon2018robust,berrone2018advanced,Flemisch2016,fum,ODS,lagrange,manzoor2018interior,sandve2012efficient,Schwenck2015}. 
The distinguishing feature between the categories is the relative fidelity with which the rock matrix and embedded fractures are represented.
Fractures are mechanical breaks distinguished from the surrounding rock matrix by the following two fundamental properties: 
first, there is a stark contrast between their length and their width.
The latter of these features, which we shall refer to as aperture, is typically much smaller than the former, i.e., a meter-long fracture can have an aperture of about $10^{-4}$ m depending on the rock type~\cite{SKB2010}.
Second, the hydraulic resistance offered by a fracture can be quite different from the surrounding rock matrix.
In turn, fractures can be hydraulic conductors, i.e., higher permeability than the rock matrix, or hydraulic barriers, i.e., lower permeability than the rock matrix.

On one end of the spectrum are continuum models where the fracture network properties are upscaled into an effective continuum and equivalent properties, e.g., bulk porosity and permeability, to represent the resistance offered by both the fracture network and matrix~\cite{hadgu2017comparative,hartley,jackson,neuman2005trends,sweeney2019upscaled,tsang1996tracer}.
In this case, all features are resolved in a single spatial dimension, typically a two or three dimensional spatial domain.  
This methodology is best suited for applications where the influence of the fracture network structure is minimal, and an effective permeability captures the bulk behavior of the system.
Examples of this situation include a densely fractured medium that behaves similar to a porous media or when the rock matrix permeability offers roughly the same resistance to flow as the fractures. 
On the other end of the spectrum are DFN models where individual fractures are explicitly represented and the surrounding matrix is not considered~\cite{berrone2013pde,berrone2015parallel,cacas1990modeling,davy2013model,davy2010likely,de2004influence,dershowitz1999derivation,dreuzy2012influence,erhel2009flow,hyman2014conforming,hyman2015dfnworks,pichot2012generalized,mustapha2007new}.
In the DFN methodology, the stark contrast in length scales of fractures leads to them being represented as $N-1$ dimensional objects embedded within an $N$ dimensional space, i.e., lines in two dimensions and planes in three dimensions.
This methodology is best suited for applications where the influence of the fracture network structure is the primary control of flow and transport behavior, and short time-scale interactions with the surrounding matrix can be neglected. 
An example of this is a sparsely fractured medium with little to no flow in the surrounding matrix.
In between the two methodologies is the discrete fracture matrix (DFM) model where both the fracture network and surrounding matrix are explicitly resolved~\cite{ahmed2015control,ahmed2015three,antonietti2016mimetic,berrone2018advanced,Flemisch2016,fum,ODS,lagrange,manzoor2018interior,Schwenck2015}. 
Similar to DFN, the fractures are represented as $N-1$ dimensional objects, but the matrix is $N$ dimensional volume surrounding the fracture network that is also considered. 
DFM models are well suited for situations where the permeability of the matrix is sufficiently high that it cannot be ignored, and the structure of the fracture network plays a key role in determining the flow field, e.g., sandstones or sealed faults.
For example, the case of geothermal systems, where flow is primarily through fractures and the rock matrix is commonly low permeability, is an application where DFMs are well suited because the quantity that \emph{flows} is heat/energy between the two domains.
This particular phenomenon of heat/energy transfer cannot be well represented by continuum or DFN models.

DFM models can be thought of as an extension of the DFN methodology where a DFN is first constructed, and then a mesh of the volume surrounding the network is generated.
In the DFN methodology, each fracture within the network is assigned a shape, location, and orientation within the domain by sampling distributions whose parameters are determined by a site characterization.
However, the size of the domains of interest, typically kilometers, and the cost of sufficiently sampling relevant quantities in the subsurface, both hydraulic and structural, results in a limited amount of data regarding the topological and geometry attributes of the fracture networks~\cite{bonnet2001scaling,national1996rock,zimmerman1993numerical}. 
In practice, these fracture networks are constructed stochastically to sample over the parameter space. 
This stochastic generation of a fracture network results in a wide range of feature length scales, which can be arbitrarily small if the generation is unconstrained.
For these features to be resolved, the local mesh resolution must be smaller than the smallest length-scale,  which is computationally impractical for arbitrarily small features.
In turn, the automated mesh generation of an arbitrary stochastically generated DFM is relatively infeasible for an unconstrained three-dimensional network composed of many fractures.

Available DFM models can be loosely partitioned into two categories based on how they address the complications associated with mesh generation. 
DFM models either have a mesh representation of the matrix that aligns/coincides with the mesh of the fracture network, referred to as a \emph{conforming mesh method}~\cite{ahmed2017cvd,bogdanov2003two,helmig1997multiphase,karimi2003efficient}, or does not, referred to as a \emph{non-conforming mesh method}~\cite{berrone2018advanced,berrone2017flow,Flemisch2016,frih2012modeling,fum,Schwenck2015}.
This computational geometry choice directs the selection or development of a numerical scheme to resolve flow in coupled fracture/matrix systems~\cite{flemisch2018benchmarks}, which is complicated by the mixed-dimensional nature of the DFM, where fractures are co-dimension 1 of the matrix.
Non-conforming methods bypass the issues associated with mesh generation.
In these methods, neither the mesh of the fracture network needs to conform to intersections between fractures, nor the mesh of the matrix needs to conform to the mesh of the fracture network.
Although mesh generation is simpler than when using a conforming method, the computational cost is pushed upward via additional degrees of freedom in the resulting discretized systems. 
To discretize the governing equations for flow and transport in the DFM, non-conforming methods have applied extended finite elements~\cite{del2017well,Flemisch2016,fum,Schwenck2015}, embedded finite elements~\cite{ODS}, finite elements/boundary elements~\cite{berrone2018advanced}, mixed finite element~\cite{boon2018robust}, mortar-type methods~\cite{frih2012modeling,nordbotten2019unified}, immersed boundaries~\cite{berrone2017flow}, and Lagrange multipliers~\cite{lagrange}.
However, the inclusion of additional physical processes, such as multiphase flow or chemical transport and reactions, commonly requires additional novel mixed-dimensional discretizations for their respective governing equations rather than more straightforward extensions of their unidimensional counterparts.
In contrast, the numerical methods for resolving flow and transport in the coupled system when using a conforming mesh are typically simpler. 
They can have fewer degrees of freedom compared to non-conforming mesh methods.
The most commonly used conservative discretization methods for conforming unstructured meshes are control-volume distributed multi-point flux approximations~\cite{ahmed2015control,ahmed2015three,ahmed2017cvd,manzoor2018interior,sandve2012efficient}, mixed finite elements,
and recently mimetic finite difference~\cite{antonietti2016mimetic}. 
These methods are widely used in non-DFM applications and have seen extensive development, including the construction of higher-order methods for flow and transport and multiphase flow, which is promising for their eventual application to DFM models~\cite{abushaikha2020fully,lipnikov2014mimetic}.
However, the trade-off is that the challenges associated with conforming mesh generation must be addressed.
While it is not required that the mesh of the fracture network conforms to intersections between fractures, it considerably reduces the complexity of the discretized governing equations and volume mesh generation when compared to non-conforming methods. 
To date, most conforming DFM models in the literature are either two-dimensional~\cite{ahmed2015control,ahmed2017cvd}, quasi two-dimensional~\cite{ahmed2015three}, or relatively simple three-dimensional geometries~\cite{berre2018flow} due to the complexity of mesh generation.

We present a conforming mesh methodology for resolving three-dimensional flow and transport that addresses the key challenges of the DFM approach in a unified manner via (1) a method for mesh generation of and around a complicated 3D fracture network, (2) the discretization of the governing equations for flow and transport using second-order accurate mimetic finite differencing, and (3) implementation of numerical methods for high-performance computing environments. 
In section~\ref{sec:equations}, we present the governing equations along with our adopted discretization method and the methodology for the mixed dimensional coupling.
We then discuss our hierarchical implementation of physical process kernels that allows for both strong and weak coupling of the fracture/matrix domains.
In section~\ref{sec:meshing}, we address the issues of conforming mesh generation by extending the feature rejection algorithm for meshing (FRAM) presented in Hyman et al.~\cite{hyman2014conforming} to produce a conforming Delaunay tetrahedralization of the volume surrounding the fracture network where the triangular cells of the fracture mesh are faces in the volume mesh. 
In section~\ref{sec:examples}, we provide two expositions of the method in complex fracture networks. 
In the first example, we demonstrate that the method is robust by considering three scenarios where the fracture is a barrier, a primary pathway, and offers the same resistance as the surrounding matrix. 
In the second example, flow and transport through a 3D stochastically generated network containing 257 fractures is presented.
General remarks and discussion about the method are provided in the final  section~\ref{sec:remarks}.

\section{Numerical Methods: Governing Equations, Discretization Methods, and Implementation}\label{sec:equations}

In this section, we present the governing equations for flow and transport in two and three dimensions and their mixed-dimensional coupling.
We then present the conservative discretization schemes and conclude with their computational implementation.

\subsection{Governing Partial Differential Equations}

The domain of interest is a fractured network surrounding by a porous media.
We denote the domain as $\Omega \subset \mathbb{R}^3$.
Our focus is on the laminar flow of an isothermal, incompressible, single-phase fluid passing through a porous media that can be described using Darcy's Law.
For simplicity, we present the equations for the case of steady flow.
Under this model description, the partial differential equations (PDE) governing flow in $\Omega$ are 
\begin{linenomath*}
\begin{equation}
\label{eq:equi-flow}
\begin{array}{rcll}
  -\nabla \cdot (\rho \bq) &=& Q, \\[1ex]
  \bq &=& -\Frac{\bK}{\mu} (\nabla p - \rho \bg), \qquad \mbox{(Darcy's law)}\\[2ex]
\end{array}
\end{equation}
\end{linenomath*}
where $\bq$ is the Darcy flux [m/s], $p$ is the aqueous pressure [Pa], $\bK_m$ is the absolute permeability tensor [m$^2$], $\mu$ is the fluid viscosity [Pa $\cdot$ s], $\rho $ is fluid density [kg/m$^3$], $\bg$ is the gravity vector, and $Q$ denotes external sources. 
The conceptual PDE model of the advective transport of non-reactive solute $C$ in $\Omega$ is
\begin{linenomath*}
\begin{equation} 
\label{eq:equi-transport}
\begin{array}{rcl}
  \Frac{\partial(\phi C)}{\partial t} &=& 
    -\nabla \cdot (\bq C),\\[1ex]
\end{array}
\end{equation}
\end{linenomath*}
where $\phi$ is the porosity [-].
Equations~\eqref{eq:equi-flow} and \eqref{eq:equi-transport} along with appropriate initial and boundary conditions describe flow and transport in $\Omega$. 

We are interested in solving~\eqref{eq:equi-flow} and \eqref{eq:equi-transport} when there are fractures present within $\Omega$. 
Due to the small size of fracture apertures relative to their length, we shall model fractures as lower dimensional objects. 
We will represent each fracture as a two-dimensional manifold in $\mathbb{R}^3$. 
We adopt a convention of subscripts $m$ and $f$ corresponding to the matrix and fracture networks, respectively.
We denote the union of all fractures in the domain as $\Omega_f$ and the surrounding rock matrix as $\Omega_m \subset \mathbb{R}^3$.
We assume that $\Omega_f$ and $\Omega_m$ are well-defined such that they make up the entire domain ($\Omega = \Omega_f \cup \Omega_m$) and that they are disjoint ($\Omega_f \cap \Omega_m = \emptyset$).
We denote the surface of $\Omega_{m}$ that is in contact with $\Omega_{f}$ as $\Gamma_{mf}$. 
Note that both $\Omega_f$  and $\Omega_m$ can be disconnected sets. 

Within $\Omega_m$, the mass balance equation and the constitutive law are similar to~\eqref{eq:equi-flow}:
\begin{linenomath*}
\begin{equation}
\label{eq:matrix-flow}
\begin{array}{rcll}
  -\nabla \cdot (\rho \bq_m) &=& Q_m, \\[1ex]
  \bq_m &=& -\Frac{\bK_m}{\mu} (\nabla p_m - \rho \bg), \\[2ex]
\end{array}
\end{equation}
\end{linenomath*}
Within $\Omega_f$, the mass balance equation and the constitutive law are given by:
\begin{linenomath*}
\begin{equation}
\label{eq:fracture-flow}
\begin{array}{rcll}
  -\nabla \cdot (\rho \bq_f) &=& -\rho \big[ \tilde{\bq}_m \cdot \bn_f \big],\\[1ex]
  \bq_f &=& -\Frac{K_f}{\mu} (\nabla p_f - \rho \bg), \\[2ex]
  \end{array}
\end{equation}
\end{linenomath*}
where $\bn_f$ is the unit normal vector pointing into the fracture domain, $K_f$ is the absolute permeability of the fracture [m$^2$],
and $\tilde{\bq}_m$ is the flux between the matrix and fracture domains.
The square brackets on the right-hand side denote the difference between values on two opposite sides of a fracture. 
Thus, the right-hand side contribution represents the mass of water going into/out of the fracture.
We impose conventional boundary conditions for both matrix and fracture domains to close the problem.
On the matrix-fracture interface the mass conservation law leads to the velocity continuity condition:
\begin{linenomath*}
\begin{equation}
  \bq_m \cdot \bn_f = \widetilde{\bq}_m \cdot \bn_f.
\end{equation}
\end{linenomath*}
In the fracture, averaging the balance equation in normal direction leads to
\begin{linenomath*}
\begin{equation}
  \big[\widetilde{\bq}_m \cdot \bn_f \big]
  = -\frac{4K_f}{\mu\, a_f} (p_f - \{p_m\})
  \equiv -k (p_f - \{p_m\}),
\end{equation}
\end{linenomath*}
where $a_f$ is the fracture aperture, and $\{p_m\}$ is the mean value of pressure in matrix 
on two sides of the fracture.
A detailed derivation could be found in \cite{jaffre-coupled}.

Similarly, the conceptual PDE model of the coupled transport of non-reactive solute is
\begin{linenomath*}
\begin{equation}
\label{eq:multi-transport}
\begin{array}{rcl}
  \Frac{\partial(\phi_m C_m)}{\partial t} &=& 
    -\nabla \cdot (\bq_m C_m),\\[1ex]
  \Frac{\partial(a_f\,\phi_f\, C_f)}{\partial t} &=& 
    -\nabla \cdot (\bq_f C_f)
    + \big[ \tilde{C} (\bq_m \cdot \bn_f) \big],
\end{array}
\end{equation}
\end{linenomath*}
subject to the Dirichlet boundary conditions on the inflow part of 
the computational domain.
$\tilde{C}$ is concentration in fracture for outflow ($\bq_m \cdot \bn_f > 0$) and in matrix for inflow.

For the purpose of this work, we drop external sources and assume that the flow is driven by boundary conditions.
Additional physical models supported by our open-source simulator include transport dispersion,  energy transport, rock and fluid compressibility, and chemical reactions, but their implementation is not described here.

\subsection{Discretization method}

We discretize the system of equations (\ref{eq:matrix-flow})-(\ref{eq:multi-transport}) using the mimetic finite difference (MFD) method~\cite{lipnikov2014mimetic}.
The MFD method combines flexibility of the finite volume and finite difference frameworks with the theoretical basis of the finite element framework. 
It allows us to use general unstructured polytopal, polygonal (2D) or polyhedral (3D), meshes to model complex geometric structures that appear in fractured porous media. 
We apply this method in both $\Omega_m$ and $\Omega_f$ domains.

\begin{figure}[htb!] \centerline{
 \begin{tabular}{cc}
\includegraphics[width=0.95\textwidth]{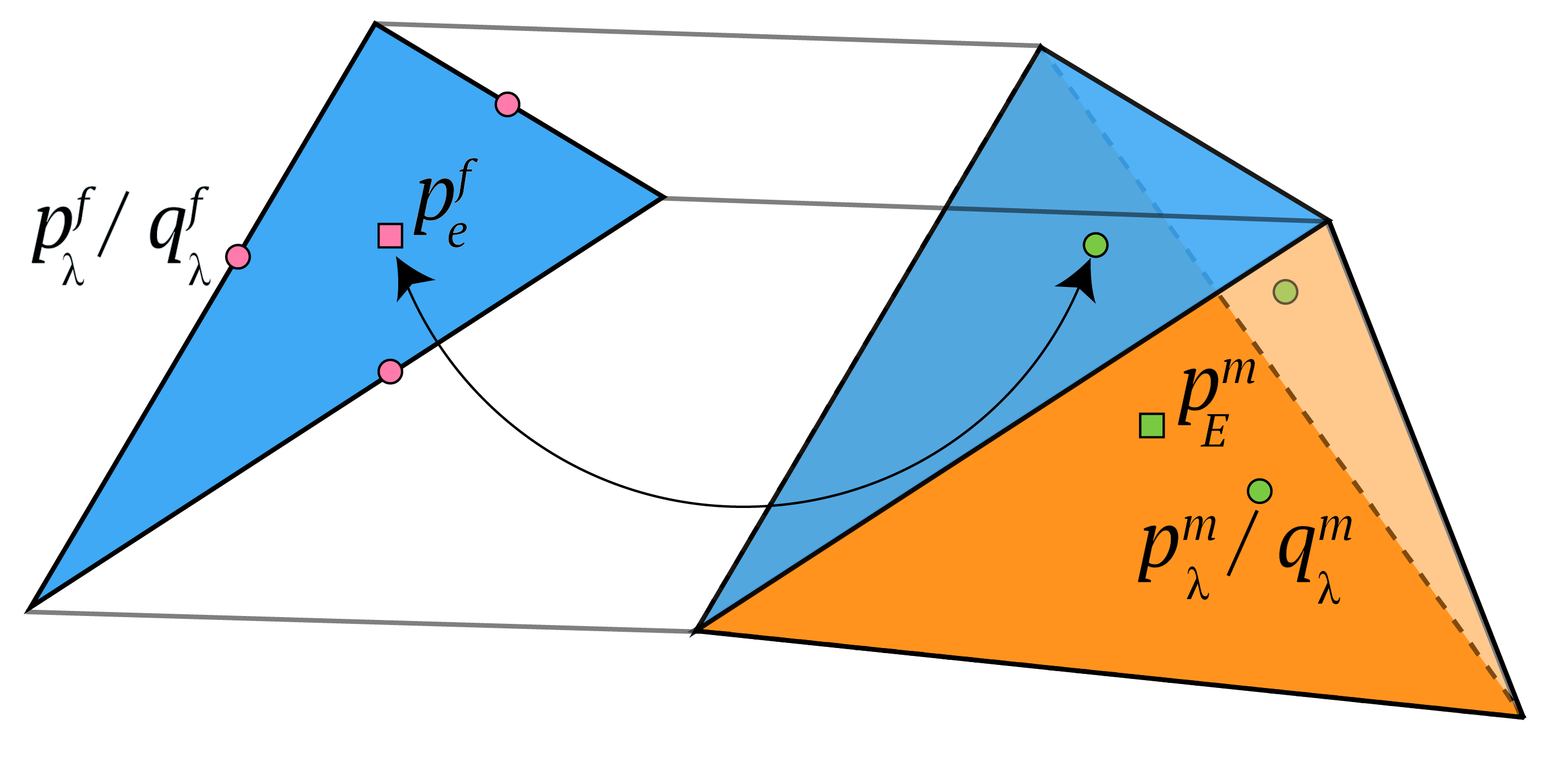} 
 \end{tabular}}
\caption{\label{fig:dfm_elements} Diagram with the positions of the unknowns on a matrix cell (tetrahedron right) and a fracture cell (triangle left). Note the the pressure unknown at the center of the fracture face is collocated with the unknown of pressure on one of the matrix cell faces. $p_E^m$: Pressure at the center of matrix cell.  $p_e^f$: Pressure at the center of fracture cell. $p_\lambda^m$: Pressure on the boundary of the matrix cell. $p_\lambda^f$: Pressure on the boundary of the fracture cell. $q_\lambda^{m}$: Normal components of the flux on boundaries of the matrix cell.
$q_\lambda^{f}$: Normal components of the flux on boundaries of the fracture cell. }
\end{figure}

To guarantee the local mass conservation property, the MFD method uses the mixed formulation and explicitly discretizes pressure and flux field unknowns.
The MFD method operates with two types of pressure unknowns: (1) cell-based, denoted with a subscript $E$ in matrix and $e$ in fractures residing at cell centers ($p_E^m$, $p_e^f$), and (2) face-based, denoted with a subscript $\lambda$ residing at cell faces ($p_\lambda^m$ and $p_\lambda^f$).
In $\Omega_m$, there is one cell-based pressure unknown $p_E^m$ per volume cell and one face-based value $p_\lambda^m$ per cell face.
In $\Omega_f$, there is one cell-based pressure unknown  $p_e^f$ per fracture cell, and one face-based value $p_\lambda^f$ per fracture cell face. 
Cells of a fracture mesh in $\Gamma_{mf}$ coincide with mesh faces  of $\Omega_m$ partition. To be able to approximate a pressure jump between two sides of a fracture surface MFD
uses two face-based pressure unknowns, $p_\lambda^{m\pm}$ that approximates pressure from both sides of a fracture. 
The flux field is expressed in terms of normal components of the flux,
$q_\lambda^{f}$ and $q_\lambda^{m}$,  which are collocated with the $p_\lambda^m$ and $p_\lambda^f$ unknowns, correspondingly.
Similar to the two pressure degrees of freedom placed on both sides of fracture, we introduce two flux unknowns $q_\lambda^{m\pm}$ as well.
Figure~\ref{fig:dfm_elements} provides a diagram with the positions of the unknowns on a matrix cell (tetrahedron right) and a fracture cell (triangle left). 

The discrete equations for the flow equations (\ref{eq:matrix-flow})-(\ref{eq:fracture-flow}) can be written locally for each matrix and fracture cell ($E$ and $e$, respectively) and then coupled and assembled into the global system. 
The local discrete equations have the following saddle-point forms:
\begin{linenomath*}
\begin{equation}
\label{eq:dis-matrix-flow}
\begin{array}{lclclcl}
{\mathbb M_E} \, \bq^m_\lambda & - & {\mathbb B}_E \,  p_E^m & + & {\mathbb C_E} \, {\bf p}_{\lambda }^m & = & 0,\\
{\mathbb B_E}^T  \bq^m_\lambda &  &  &  &  & = & Q_{m,E},
\end{array}
\end{equation}
\end{linenomath*}
and
\begin{linenomath*}
\begin{equation}
\label{eq:dis-fracture-flow}
\begin{array}{lclclcl}
{\mathbb M_e} \, \bq^f_\lambda & - & {\mathbb B}_e \,  p_e^f & + & {\mathbb C_e} \, {\bf p}_{\lambda }^f & = & 0,\\
{\mathbb B_e}^T  \bq^f_\lambda &  &  &  &  & = & Q_e^{m/f}.
\end{array}
\end{equation}
\end{linenomath*}
where $\bq^m_\lambda$,$\bq^f_\lambda$, ${\bf p}_{\lambda }^m$, ${\bf p}_{\lambda }^f$  are vectors of degrees of freedom
which correspond to a particular matrix cell, $E$, or fracture cell, $e$. 
The construction of the square mass matrices ${\mathbb M}_{E}$ and ${\mathbb M}_e$ is the key ingredient of the MFD construction, which is explained in detail in~\cite{lipnikov2014mimetic}. 
These matrices are constructed to guarantee the second order of accuracy of a numerical solution on unstructured polyhedral meshes with a full anisotropic permeability tensor.
$Q^{m/f}$ represents fluxes between $\Omega_m$ and $\Omega_f$:
\begin{linenomath*}
\begin{equation}
\label{interface-term1}
  Q_e^{m/f} = k(p_\lambda^{m,+} - p_e^f) + k(p_\lambda^{m,-} - p_e^f).
\end{equation}
\end{linenomath*}
The local systems are coupled by the flux continuity conditions inside $\Omega_m$ and $\Omega_f$ and on $\Gamma_{mf}$:
\begin{linenomath*}
\begin{equation}
  \label{vel-continuity}
  q_\lambda^{m,-}  = k(p_\lambda^{m,-} - p_e^f),
  \quad
  q_\lambda^{m,+}  = k(p_\lambda^{m,+} - p_e^f)
\end{equation}
\end{linenomath*}
From an implementation viewpoint, it is convenient to treat flux continuity conditions (\ref{vel-continuity}) as the flux boundary conditions for the matrix domain.
This leads to uniform code implementation in interior and near-boundary mesh cells.
Recall, that $\pm$ denote face-based degrees of freedom on the opposite sides of fracture cell $e$.
The last formula shows the convenience of using face-based pressure unknowns in the numerical scheme.
The coupling with the fracture domain and the approximation of a pressure jump is included into the flux continuity conditions (\ref{vel-continuity}).

Using the static condensation algorithm \cite{MFD2006} the flux degrees of freedom can be excluded locally which converts local systems
from the saddle-point form to symmetric positive definite (SPD) formulations:
\begin{linenomath*}
\begin{equation}
\label{eq:pl-matrix-flow}
\begin{array}{lclcl}
{\mathbb S}^p_{E}  \, p_E^m & + & {\mathbb S}^{p\lambda}_E \, {\bf p}_{\lambda }^m & = & Q_{m,E},\\
{\mathbb S}^{{\lambda}p}_{E}  \, p_E^m  &+& {\mathbb S}^{\lambda}_E \, {\bf p}_{\lambda }^m & = & -\tilde Q_e^{m/f},\\
\end{array}
\end{equation}
\end{linenomath*}
and
\begin{linenomath*}
\begin{equation}
\label{eq:pl-fracture-flow}
\begin{array}{lclcl}
{\mathbb S}^p_{e}  \, p_e^f & + & {\mathbb S}^{p\lambda}_e \, {\bf p}_{\lambda }^f & = & 0,\\
{\mathbb S}^{{\lambda}p}_{E}  \, p_e^f  &+& {\mathbb S}^{\lambda}_e \, {\bf p}_{\lambda }^f & = & Q_e^{m/f},\\
\end{array}
\end{equation}
\end{linenomath*}
where entries of $\tilde Q_e^{m/f}$ are either zeros or one of the terms in $Q_e^{m/f}$.

Equations (\ref{interface-term1})-(\ref{vel-continuity}) can be assembled into the global 
coupled system with a symmetric positive definite matrix:
\begin{linenomath*}
\begin{equation}
\label{eq:coupled-flow}
\left(
\begin{array}{cccc}
{\mathbb S}^p_m & {\mathbb S}^{p\lambda}_m & 0&0\\[1ex]
{\mathbb S}^{{\lambda}p}_m & {\mathbb S}^\lambda_m &{\mathbb S}_{m/f} &0\\[1ex]
0&{\mathbb S}_{m/f}^T&{\mathbb S}^p_m & {\mathbb S}^{p\lambda}_m \\[1ex]
0&0&{\mathbb S}^{{\lambda}p}_m & {\mathbb S}^\lambda_m
\end{array}
\right)
\left(
\begin{array}{c}
{\bf P}^m\\[1ex]
{\bf P}_{\lambda }^m\\[1ex]
{\bf P}^f\\[1ex]
{\bf P}_{\lambda }^f
\end{array}
\right)
=
\left(
\begin{array}{c}
\bQ_m\\[1ex]
0\\[1ex]
0\\[1ex]
0
\end{array}
\right),
\end{equation}
\end{linenomath*}
where the ${\bf P}$-vectors collect all respected local pressure unknowns.
The final step is to apply the Dirichlet boundary conditions.
This means to prescribe known values to  $p^m_\lambda$ and $p^f_\lambda$ 
located on the domain boundary and to eliminate them from the global system.


To discretize the system of hyperbolic transport PDEs \eqref{eq:multi-transport}, we apply the finite volume method. 
The solute concentration is discretized by cell-centered degrees of freedom for volumetric matrix cells and surface fracture cells.
The advective solute fluxes are approximated using the first-order upwind method (see, e.g. \cite{Barth:91}).
The solute flux is defined similarly in $\Omega_m$ and $\Omega_f$.
Consider $\Omega_m$ and some mesh face $f$ with the corresponding Darcy flux $q$ which is one of the $q^m_\lambda$ defined above.
Then, the solute flux is approximated using the upwind scheme:
\begin{linenomath*}
\begin{equation}\label{solute_flux}
  \int_{f} C_m (\bq \cdot \bn)\, {\rm d}S \approx |f|\, (q^{+} C^m_{E,+} + q^{-} C^m_{E,-}),
\end{equation}
\end{linenomath*}
where
\begin{linenomath*}
\begin{equation}
  q^{+} = \frac{1}{2}(q^* + |q^*|), \qquad q^{-} = \frac{1}{2}(q^* - |q^*|),
\end{equation}
\end{linenomath*}
and $C^m_{E,\pm}$ are solute concentrations in the cells on both side of a face $f$.
Here, we assume that the normal $\bn$ points from cell $+$ to cell $-$.
We denote the right-hand side of equation \eqref{solute_flux} as $|f|\,(q\,C)^m_\lambda$
by analogy with the flux definition.

Using the first-order time approximation, we replace equations \eqref{eq:multi-transport}
with discrete equations
\begin{linenomath*}
\begin{equation}
\label{eq:multi-transport-discrete}
\begin{array}{rcl}
  \Frac{\phi_m (C_E^{m,n+1} - C_E^{m, n})}{\Delta t} &=& 
    -\displaystyle\sum\limits_{f \in \partial E} |f| (q\,C)^{m, n+\theta}_\lambda,\\[1ex]
  \Frac{a_f\,\phi_f\, (C_e^{f, n+1} - C_e^{f, n})}{\Delta t} &=& 
    -\displaystyle\sum\limits_{g \in \partial e} |g|\, (q C)_\lambda^{f, n+\theta}
    + \big[ \tilde{C} (\bq_m \cdot \bn_f) \big].
\end{array}
\end{equation}
\end{linenomath*}
Setting $\theta=0$, we obtain the explicit scheme. 
The implicit scheme is obtained by setting $\theta=1$.
A second-order advection scheme is derived within the simulator using more accurate approximation of the solute flux based on a limited linear reconstruction of the concentration. A second-order  accurate time integration is implemented via the Runge-Kutta scheme.

\subsection{Implementation: Process Kernels}

We implemented the coupled matrix-fracture flow system in the open-source code Amanzi~\cite{ASCEM_Amanzi_design_1} which provides a suite of critical low-level components and is built on the flexible and extensible Arcos process management system~\cite{ETCoon_JDMoulton_SLPainter_2016a}. 
Amanzi provides a flexible parallel mesh infrastructure that supports the extraction of meshes on the matrix and fracture sub-domains, i.e., $\Omega_f$ and
$\Omega_m$.  
It provides an operators library that contains MFD and FV discretizations of the flow and transport equations. 
In addition, it includes a suite of nonlinear and linear solvers \cite{KLipnikov_DMoulton_DSvyatskiy_2016a}, and provides interfaces to powerful third-party solvers such as Hypre's algebraic multigrid.

\begin{figure}[htb!]
  \begin{center}
    \includegraphics[width=0.65\textwidth]{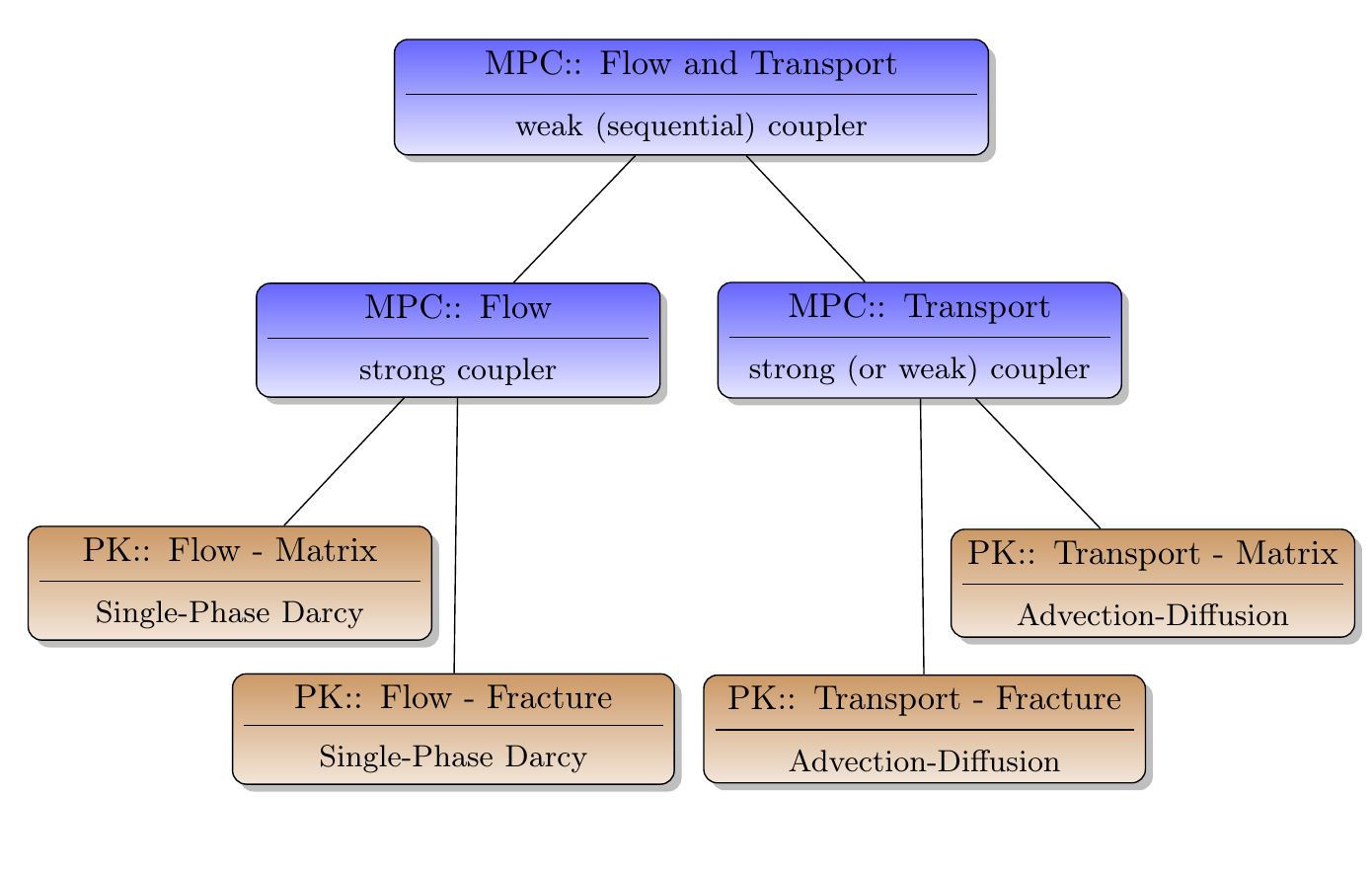}
    \caption{\label{fig:amanzi-pk-tree} The Amanzi Process Kernel (PK) tree is shown for the matrix-fracture flow and transport system. Flow and Transport are coupled sequentially. Flow System: Flow Coupler - tight coupling forms and solves matrix-fracture flow as a single system. Flow PK on matrix mesh generates sub-system. Flow PK on fracture mesh generates sub-system. Transport System: Transport Coupler: tight coupling solves matrix-fracture transport implicitly. Can sub-cycle time-steps relative to flow. Transport PK on matrix mesh generates sub-system. Transport PK on fracture mesh generates sub-system.  }
  \end{center}
\end{figure}

Arcos, the underlying framework of Amanzi, provides services to manage process-rich systems with many coupled models, including both PDE and parametric models. 
At a high level, Arcos defines two key building blocks or objects, the process kernel (PK) and the multi-process coordinator (MPC).  
The process kernel provides an implementation of an API for a single process, e.g., a single PDE, on a specific mesh. 
The MPC provides an implementation of a particular coupling strategy using this same PK API to assemble the PKs and MPCs hierarchically in a PK-tree to represent the coupled system. 
A PK-tree for the coupled matrix-fracture flow and transport system is
shown in Figure \ref{fig:amanzi-pk-tree}.  
The tight coupling of the matrix-fracture flow system shown there is realized, at each time step, by solving the global matrix formed by combining the associated matrix and fracture matrices.  
The vector and operator abstractions in Amanzi make this tight coupling relatively easy to implement.  
In contrast, the matrix-fracture transport can be weakly coupled, with implicit diffusion and explicit advective transport in the matrix and the fracture, or strongly coupled with the implicit solution of a single advection-diffusion system formed by combining the matrices from the matrix and fracture systems.
The top-level or base MPC shows that these flow and transport MPCs are coupled sequentially. 
The flow time step is usually much larger than a stable and accurate transport time step, so the transport MPC is sub-cycled relative to the flow MPC.

At a lower level, Arcos dynamically manages a complete and self-consistent view of all variables in the system, dubbed the state. 
For example, the instantiation of PK or MPC registers its data requirements, i.e., pressure on cells of the matrix mesh, with the state data manager so that Arcos can construct a directed acyclic graph (DAG) of all data dependencies in the system.  
Acros uses this DAG to manage updates of all variables in the system, eliminating code from PKs and MPCs that would depend on their relative position in a hierarchy. 

Accuracy and efficiency of Amanzi framework was successfully tested in the benchmark studies \cite{berre2021verification}. Amanzi code was one of a few participants that successfully simulated all proposed scenarios ranging from one fracture problem up to field scale scenarios based on DFM network with 52 stochastically generated fractures.

\section{Conforming Mesh Generation}\label{sec:meshing}
The discretization provided in Section~\ref{sec:equations} requires that the two-dimensional mesh of $\Omega_f$ is the trace of the three-dimensional mesh of $\Omega_m$. 
In this section, we describe our methodology to accomplish this task.  

\subsection{DFN Meshing}\label{sec:dfn_meshing}
We generate and mesh each fracture network using the Feature Rejection Algorithm for Meshing ({\sc FRAM})~\cite{hyman2014conforming} implemented within the {\sc dfnWorks} computational toolkit~\cite{hyman2015dfnworks}.
{\sc FRAM} produces a conforming Delauney triangulation of a DFN without user intervention or manual mesh manipulation by coupling the generation of the networks with meshing.
The cornerstone of {\sc FRAM} is a user-defined minimum length scale ($h$) that determines what geometric features will be represented in the network and is conceptually equivalent to the mesh resolution.
{\sc FRAM} constrains the network generation so that the local feature size on each fracture is greater than $2h$.
This constraint ensures that pathological cases that degrade mesh quality do not exist and thus provides a firm lower bound on the required resolution of the mesh.
In turn, the network can be meshed using cells with edges less than $\approx h$.
The algorithm presented by Murphy et al.,~\cite{murphy2001point} is used to produce a conforming Delaunay triangulation where contiguous edges in the mesh represent the lines of intersections between fractures.
We refer the reader to Hyman et al.~\cite{hyman2014conforming} for details concerning {\sc fram}.
While {\sc fram}  and {\sc dfnWorks} can generate meshes with either variable or uniform resolution, we only consider a uniform resolution mesh as it simplifies the extension to the creation of a conforming volume mesh.

\subsection{DFM Meshing}
The meshed DFN produced by {\sc fram} is the input for our volume meshing strategy.
The volume mesh generation workflow presented below is an automated approach to filling the volume around and between the triangulated planar polygons of the DFN with a tetrahedral mesh that conforms to the triangle facets of the DFN. 
While it is not a requirement for the discrete formulation presented in Sec.~\ref{sec:equations} the triangulations that tessellate each planar fracture polygon are Delaunay triangulations and the tetrahedral mesh that fills the volume is also a Delaunay mesh. 
The workflow described is a strategy to generate a distribution of vertices in 3D such that a Delaunay tetrahedralization of the vertices will result in a mesh with tetrahedron edges that conform to the lines of intersection between each planar fracture and a mesh with tetrahedron triangle facets that conform to the triangulation of the planar polygon fractures. 
The theoretical basis for the approach is proven in Murphy et al.~\cite{murphy2001point}. 
The method is presented for the case where the triangulated DFN has uniform resolution, the edge lengths of the triangles have similar length.
There is no theoretical barrier for the method to be applied when the mesh resolution varies from coarse to fine.

\begin{figure*}[p] \centerline{
 \begin{tabular}{cc}
\includegraphics[width=0.95\textwidth]{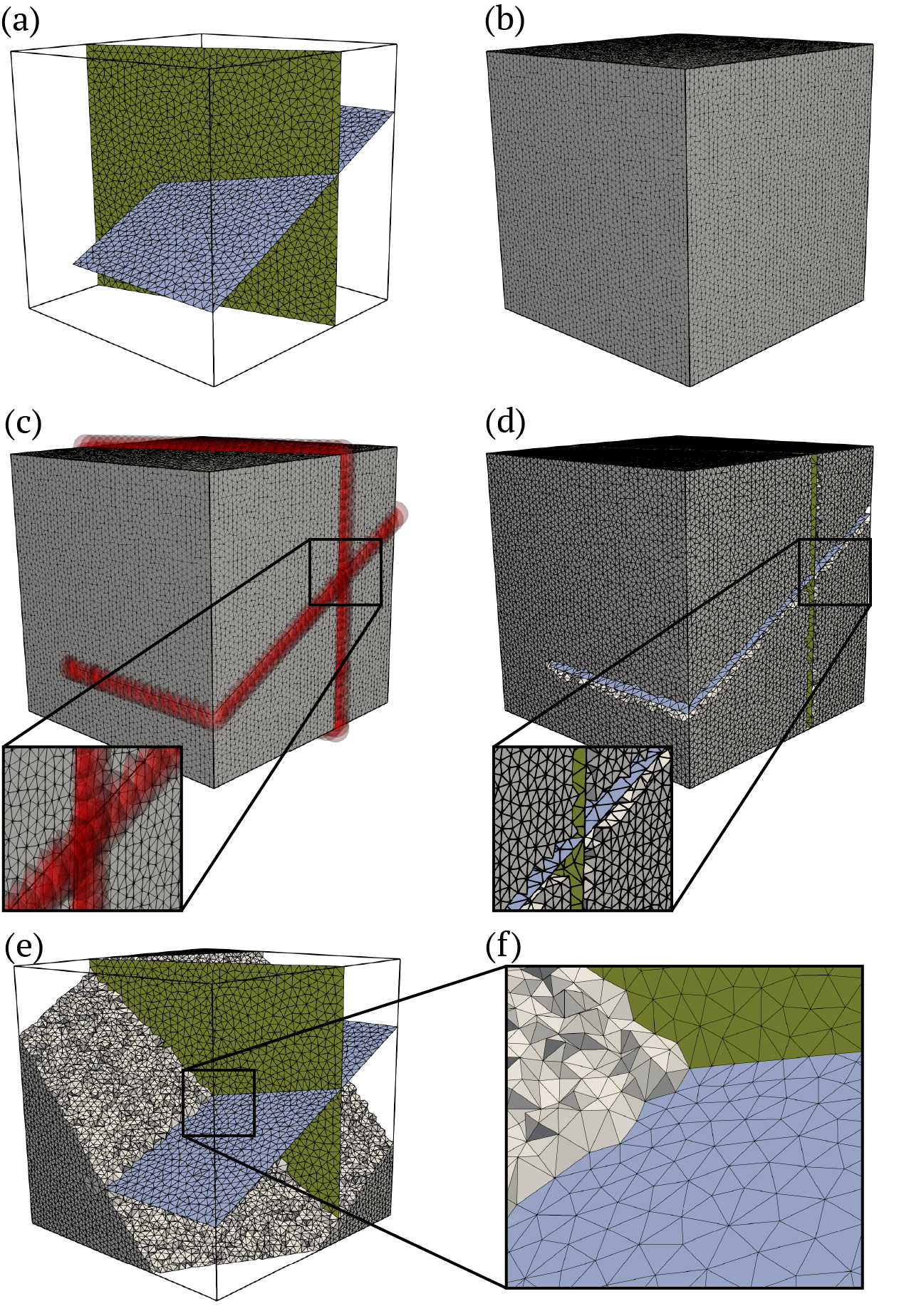} 
 \end{tabular}}
\caption{\label{fig:meshing_workflow} Visual explanation of the DFM meshing workflow. (a) DFN composed of two fractures with the Delaunay Triangulation overlaid in black. (b)  Isotropic tetrahedral mesh whose faces cells are the same resolution as the DFN mesh. (c) Red spheres with radii $h$ are placed onto every node in the mesh of DFN. (d) Tetrahedral mesh where vertices within the spheres have been removed. (e) Reconnected tetrahedral mesh where the DFN cells are now faces in the tetrahedral mesh. (f) Close up view of the hybrid DFM mesh which is a Delaunay tetrahedralization where the fracture triangulation is represented as faces of the tetrahedrons.  }
\end{figure*}

The method is broken up into three primary steps outlined below. 
Figure~\ref{fig:meshing_workflow} provides a visual depiction of the workflow and psuedo-code is provided in  Algorithm 1 presented in~\ref{app:meshing}.
We use a network composed of two rectangular fractures for visual exposition of the workflow, but the algorithm is general for an $n$ fracture DFN.

\begin{enumerate}

\item [Input-] Generate a uniform resolution conforming Delaunay triangulation of a DFN. 
Our example case of a two fracture DFN, one green and one blue, with the mesh overlaid in black is shown in Fig.~\ref{fig:meshing_workflow}~(a). 
The fractures are triangulated such that the triangulations are conforming, i.e., share edges, along the lines of intersection.
This mesh is generated using the method provided in Sec.~\ref{sec:dfn_meshing}.
Triangular edges in the mesh are of length $\approx h$. 

\item[Step 1-] Create a background mesh, which is a directionally isotropic tetrahedrons mesh with cell edge length $\approx h$, which matches the size of the DFN triangles.
We generate the background mesh by estimating the number of vertices necessary to produce a uniform sized mesh with the desired spacing and generate a distribution of vertices in a cubic pattern.
This point distribution is connected to form Delaunay tetrahedrons, which is then smoothed using elliptic smoothing that guards against inverting cells~\cite{khamayseh1996anisotropic}.
Then, a final set of smoothing steps is performed using a three-dimensional implementation of the minimum error gradient adaption algorithm that creates a smoothed grid that is adapted to the standard function with constant Hessian~\cite{bank1997mesh}.
The first smoothing step breaks the symmetry of the initial orthogonal mesh, which is a false local minimum of the smoothing algorithm, and the second smoothing step converges to a high quality, isotropic mesh. 
This methods also contains features that prevent inverting cells during vertex movement. 
However, these smoothing steps do not maintain a Delaunay mesh.
Thus, the final step is to throw away the connectivity and then reconnect the vertices using a three-dimensional Delaunay tetrahedralization algorithm~\cite{lagrit2011}.
The resulting mesh is shown in Fig.~\ref{fig:meshing_workflow}-(b).
This approach, while being effective at creating a high quality, 
isotropic, 3D Delaunay mesh, is inefficient with respect to use of computing resources~\cite{kuprat1996adaptive}.
Other methods, such as Poisson disk sampling in 3D~\cite{guo2016tetrahedral,krotz2021maximal} may prove to be a more efficient approach, but has not been implemented or utilized for the calculations presented here.

\item[Step 2-] Excavate the background mesh. 
We superimpose the triangle DFN  and the tetrahedron mesh to compute the circumscribed sphere of each triangle in the DFN mesh, which are shown in Fig.~\ref{fig:meshing_workflow}~(c).
If a vertex of the tetrahedron mesh falls within any circumscribed sphere it is removed along with any tetrahedron associated with the vertex.
We call this the \emph{excavated} mesh, which is shown in Fig.~\ref{fig:meshing_workflow}~(d).
This criteria, that every edge or triangle facet of the input DFN that must have an empty circumsphere, is a sufficient condition so that the output DFM Delaunay mesh conforms to those edges and facets~\cite{murphy2001point}.
In order to maximize mesh quality, the distance field between vertices of the DFN and the tetrahedron mesh is computed and any vertex within distance $h/2$ is also removed.
This last step is not necessary for obtaining the desired conforming Delaunay mesh, but experience has shown that the mesh quality is improved in terms of tetrahedron aspect ratio.

\item[Step 3-]  Create DFM mesh.
We create a new point cloud that is the union of the DFN vertices and remaining vertices in matrix mesh. 
Vertices in this point cloud are then connected using a three-dimensional Delaunay tetrahedralization algorithm. 
The resulting mesh is shown in Figure~\ref{fig:meshing_workflow}~(e).
These previous steps and the theory presented in Murphy et al.~\cite{murphy2001point} ensure that (i) all the lines of intersection between the fractures will exist as edges of the tetrahedron mesh and (ii) the triangles of the DFN mesh will exist as faces of the cells of the tetrahedron mesh. 
Figure~\ref{fig:meshing_workflow}~(f) is a close up of the mesh along the line of intersection between fractures to show that the tetrahedral mesh conforms to the DFN mesh as well as the line of intersection.
The final step required for setup of the physics model is to determine this correspondence between the triangles in the DFN mesh and the tetrahedron faces in the DFM, which is a matter of book keeping.
\end{enumerate}

The workflow outlined above is robust by design, especially when the DFN is sparse and the angle between the fractures is sufficiently large.
However, we acknowledge the method presented here is not perfect for any arbitrary networks.
It is possible, albeit rare (typically $< 1\%$), for few facets of the original DFN to not be maintained along the fracture intersections. 
The reason for this is that while we ensure that no vertices of the DFM are within the circumsphere of the DFN triangles, we do not ensure that a vertex from a nearby intersecting fracture triangle is not within any DFN circumsphere. 
For most physical scenarios where a DFM model would be appropriate, this is a small part of the overall coupled fracture-matrix system, and the effect on flow, transport, and other physical quantities of interest is negligible.
Nonetheless, we propose two optional extensions of methods to address these interference points in the meshing procedure. 
The first is a three-dimensional extension of {\sc FRAM} and the second is a graph-based/upscaling method that reduces the network complexity while preserving key hydrological properties of the fracture and matrix system.

\subsection{A Three-Dimensional Extension of the Feature Rejection Algorithm for Meshing}
The version of {\sc FRAM} presented in Hyman et al.~\cite{hyman2014conforming} used to create the DFN mesh does not consider features in three-dimensional space that affect the ability to automatically generate a conforming volume mesh of the surrounding matrix.
Examples of three-dimensional features that influence volume meshing, but not that of the DFN, include the distance between two fractures that do not intersect and the angle of intersection between two fractures.
Suppose two fractures are arbitrarily close but do not intersect. In that case, the volume cell between the two planes must have edge lengths smaller than that distance, which is impracticable if the minimum distance is unconstrained.

To address these issues, one can extend {\sc FRAM} to constrain the generation of the network to adhere to a minimum feature size in three-dimensions and automatically produce a conforming volume mesh using the method presented above.
While this methodology is straightforward to implement, it is computationally costly as it must be exhaustive.
If the network is sufficiently sparse and contains a few primary features, e.g., large faults, then this method will produce a network that can be automatically meshed, i.e., without user manipulation, 
However, as networks become denser, it becomes impracticable to constrain the network so that all three-dimensional features are larger than $h$ while adhering to geological observations and keeping the number of mesh cells to a reasonable number.


\subsection{Network Simplification Through Primary Network Identification}

Flow channeling, isolated regions of high velocity within the flow field, is commonly observed in field observations, laboratory experiments, and numerical simulations in fracture media~\cite{abelin1991large,abelin1985final,hyman2020flow,dreuzy2012influence,rasmuson1986radionuclide}.
These observations indicate the existence of primary sub-networks, also referred to as the network {\it backbones}, where the majority of flow and transport occurs.
These backbones are commonly composed of a few large fractures connected by small fractures, and they tend to be structurally less complicated than the entire network due to their reduced size. 
The formation of flow channels suggests that in many applications, it may be unnecessary to explicitly resolve all fractures in the network to effectively capture the system's bulk flow and transport behavior.
The method proposed below exploits this feature by partitioning a network into primary and secondary subnetworks where the DFM volume mesh conforms to the mesh of the primary network and influence of the secondary network is accounted for using upscaled effective parameters.

The conceptual definition of a backbone is relatively simple but formulating an operational definition to partition a fracture network depends on various variables.
Formally, we seek to partition$ \Omega_f$ in two disjoint subsets, the backbone $\Omega_f^\prime$ and the secondary structure $\Omega_f^\ast$ such that they make up the whole network $\Omega_f = \Omega_f^\prime \cup \Omega_f^\ast$ and they are disjoint $\Omega_f^\prime \cap \Omega_f^\ast = \emptyset$. 
Backbone membership has been shown to depend on the macro-scale structure of the network, e.g., orientations and density~\cite{hyman2019linking,sherman2020characterizing,hyman2020flow}, meso-scale hydrological attributes, e.g., fracture permeability~\cite{de2002hydraulic,hyman2016fracture}  as well as the flow domain boundary conditions~\cite{grindrod1993channeling,neuman2005trends}. 
There are a number of methods to identify networks backbones~\cite{aldrich2017analysis,hyman2017predictions,hyman2018identifying,maillot2016connectivity,wellman2009effects}
In principle, any of these methods can be applied to identify $\Omega_f^\prime$ and its compliment. 
Once the backbone is identified, we shall generate a volume mesh that only conforms to the mesh of the backbone. 
This process reduces the complexity of generating a volume mesh around the entire network because fewer small local features that lead to degradation of mesh quality due to interference from nearby points exist.
Once the backbone is identified and the matrix surrounding it meshed, we use the methods developed by Sweeney et al. \cite{sweeney2019upscaled} to upscale the properties of the secondary network $\Omega_f^{\ast}$ into equivalent properties of the matrix mesh. 
We refer the reader to the original paper for complete details, but we highlight the key aspects here.
The basis for calculating equivalent permeabilities and porosities of the secondary fractures is to calculate their intersection areas with the matrix tetrahedron. 
Once the intersection areas are known, the porosity and permeability of each intersected cell can be calculated using the fracture apertures and orientations. 
This procedure is done for each tetrahedron cell in the matrix mesh and allows for properties of multiple fractures  (intersecting fractures) to be upscaled to the same cell.

\subsubsection{Network Simplification: Example}\label{sec:backbone_example}

\begin{figure*}[htb!] \centerline{
 \begin{tabular}{cc}
\includegraphics[width=0.45\textwidth]{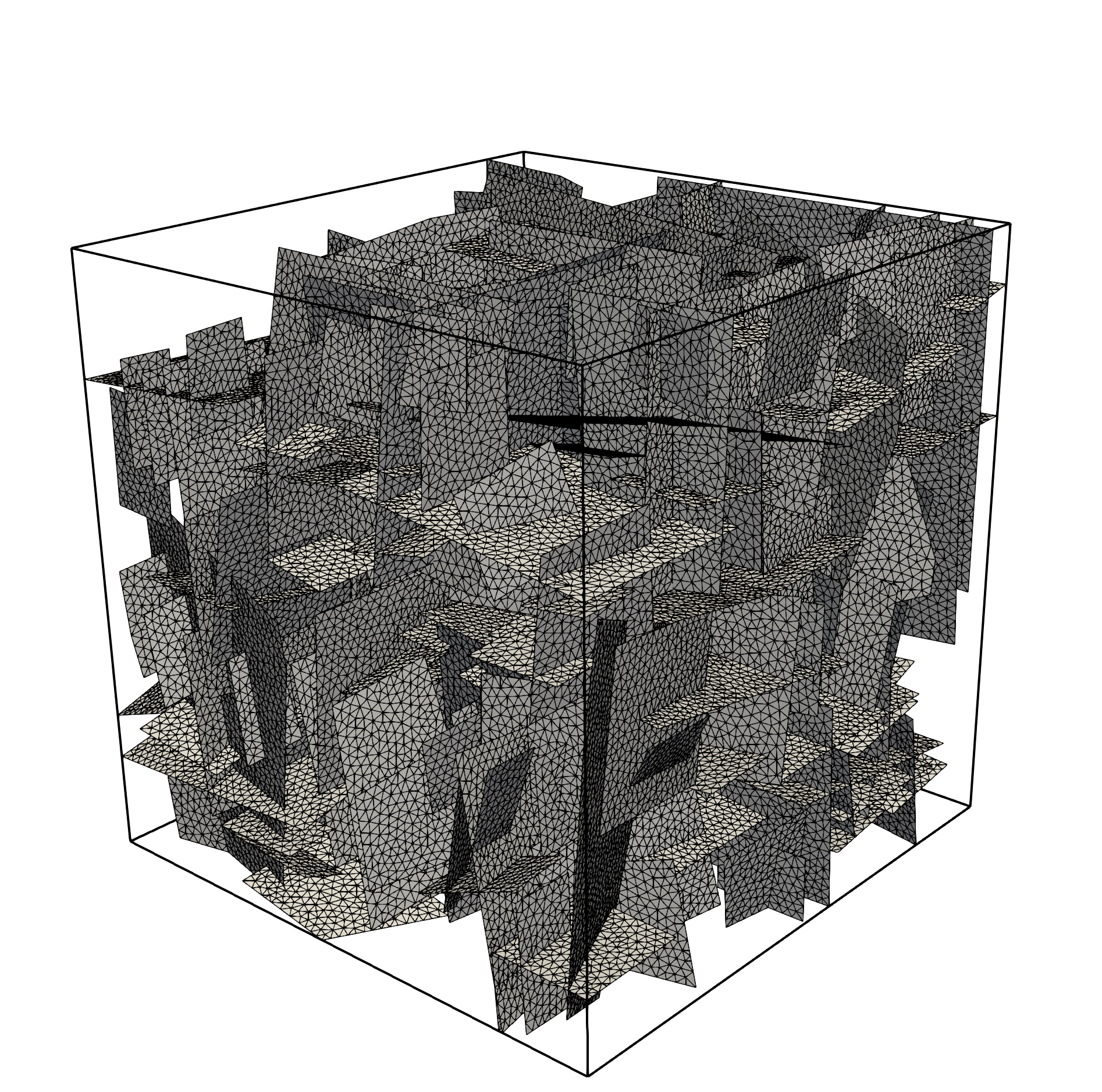} & \includegraphics[width=0.45\textwidth]{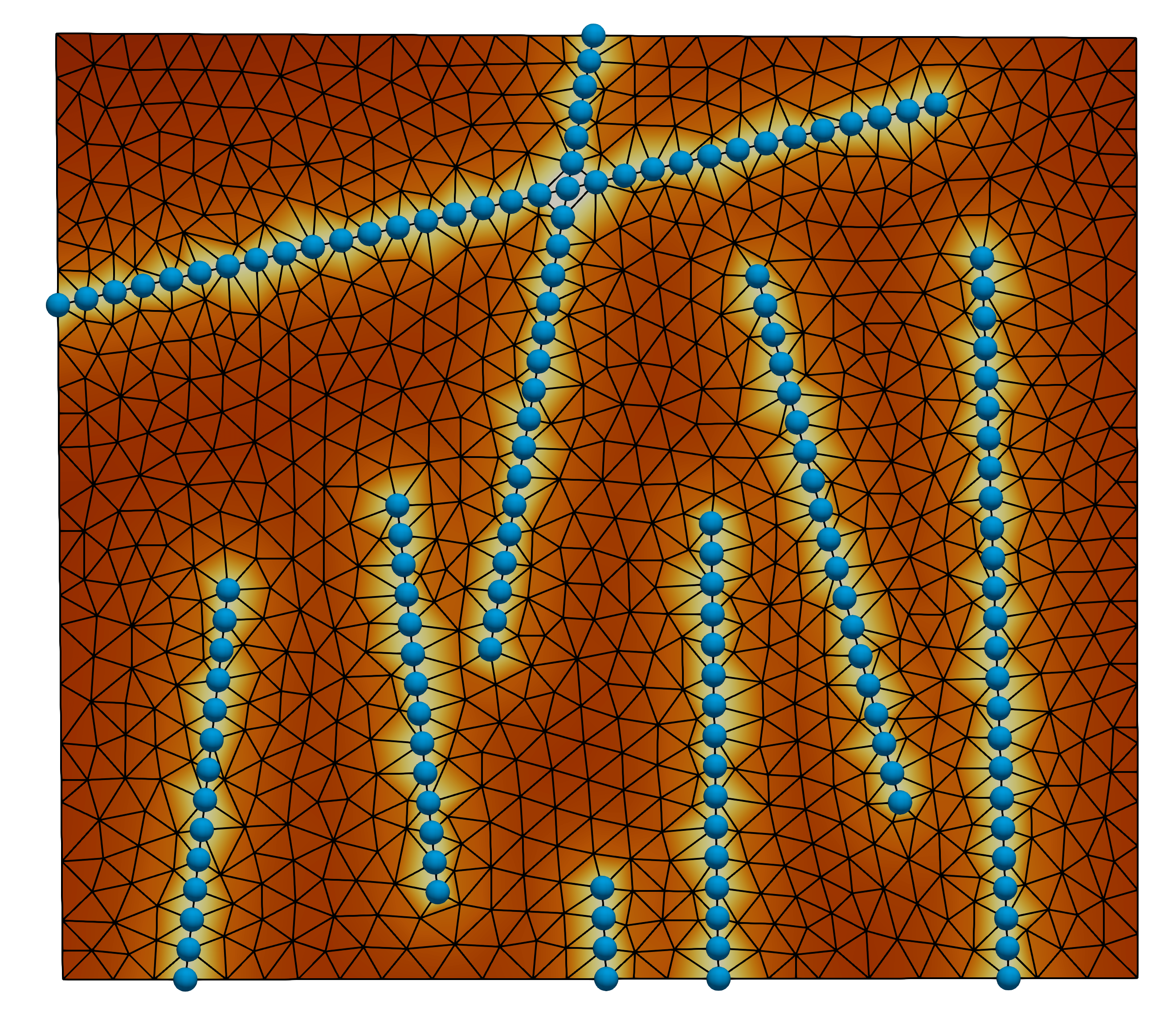}  \\
\includegraphics[width=0.45\textwidth]{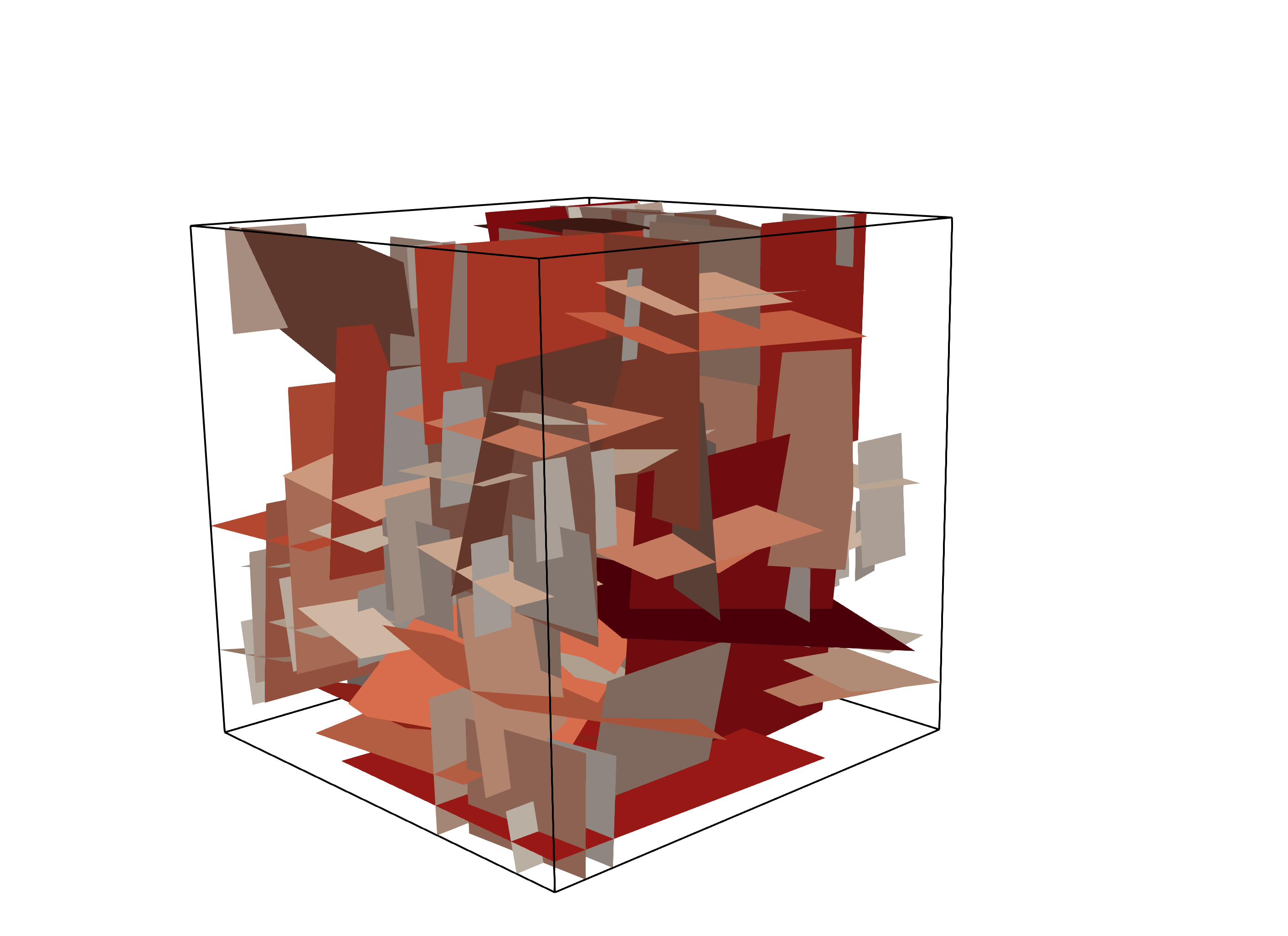} & \includegraphics[width=0.45\textwidth]{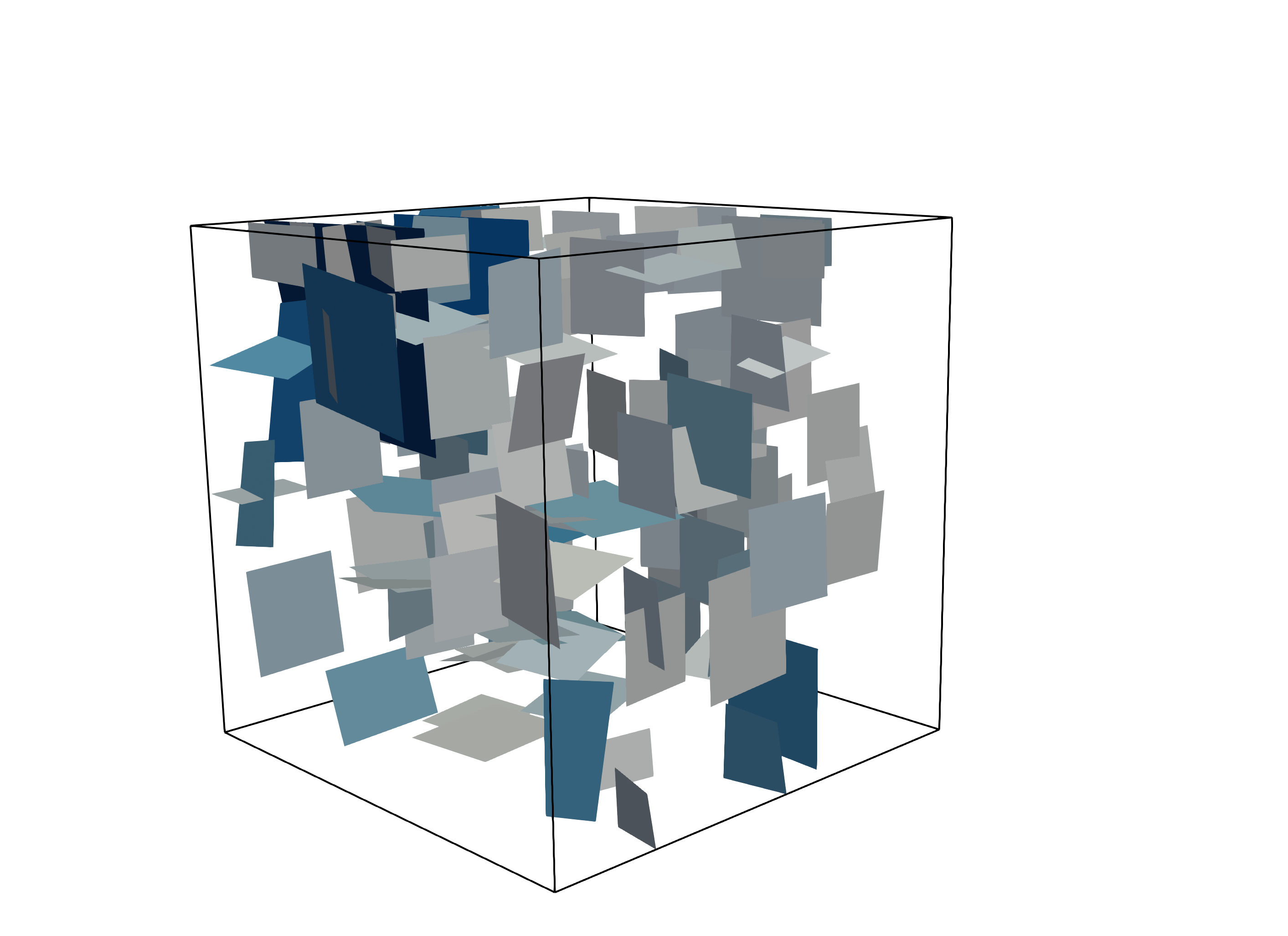} 
 \end{tabular}}
\caption{\label{fig:dfn_mesh} (Top Left) DFN composed of 257 Fractures. Mesh has 142841 cells and 74418 vertices. (Top Right) Single fracture from within network shown on left. 1790 cells and 949 vertices. (Bottom Left) Primary subnetwork/backbone. (Bottom Right) Secondary sub-network.}
\end{figure*}

Figure~\ref{fig:dfn_mesh} (top left) provides a example meshed DFN generated using {\sc fram}, which we will use to demonstrate the network simplification method.
The domain is a cube with sides of length 10 meters.
The fractures are squares whose lengths are sampled from a truncated power-law distribution with an exponent of 2.4, lower cutoff of 1.5 meters, and upper cutoff of 4.5 meters. 
The fractures in the network are sampled from three different families whose mean orientations align with the principal Cartesian axis.  
There is slight variation around the mean direction in each family by sampling the normal vectors using a von-Misser Fisher distribution with dispersion parameter $\kappa=40$~\cite{wood1994simulation}.
The resulting network is made up of 257 fractures.
In this network, $h$ is set to 0.25 m. 
Overlaid on the DFN is the conforming Delaunay triangulation, which contains 142841 triangular cells and 74418 vertices.
Figure~\ref{fig:dfn_mesh} (top right) shows the mesh of a single fracture extracted from the network. 
The mesh is isotropic and composed of uniformly size triangles. 
This particular fracture contains 1790 cells composed of 949 vertices.
Vertices along the lines of intersection are blue spheres.
Cells whose edges align with the line of intersections are yellow, and the remaining cells are colored orange. 
This particular fracture intersects eight other fractures and includes one point where there is an intersection of intersections on the fracture plane, i.e., a triple intersection.
Note that the lines of intersections, including triple intersections, are represented in the triangulation. 

\begin{figure}[htb!] \centerline{
 \begin{tabular}{cc}
\includegraphics[width=0.95\textwidth]{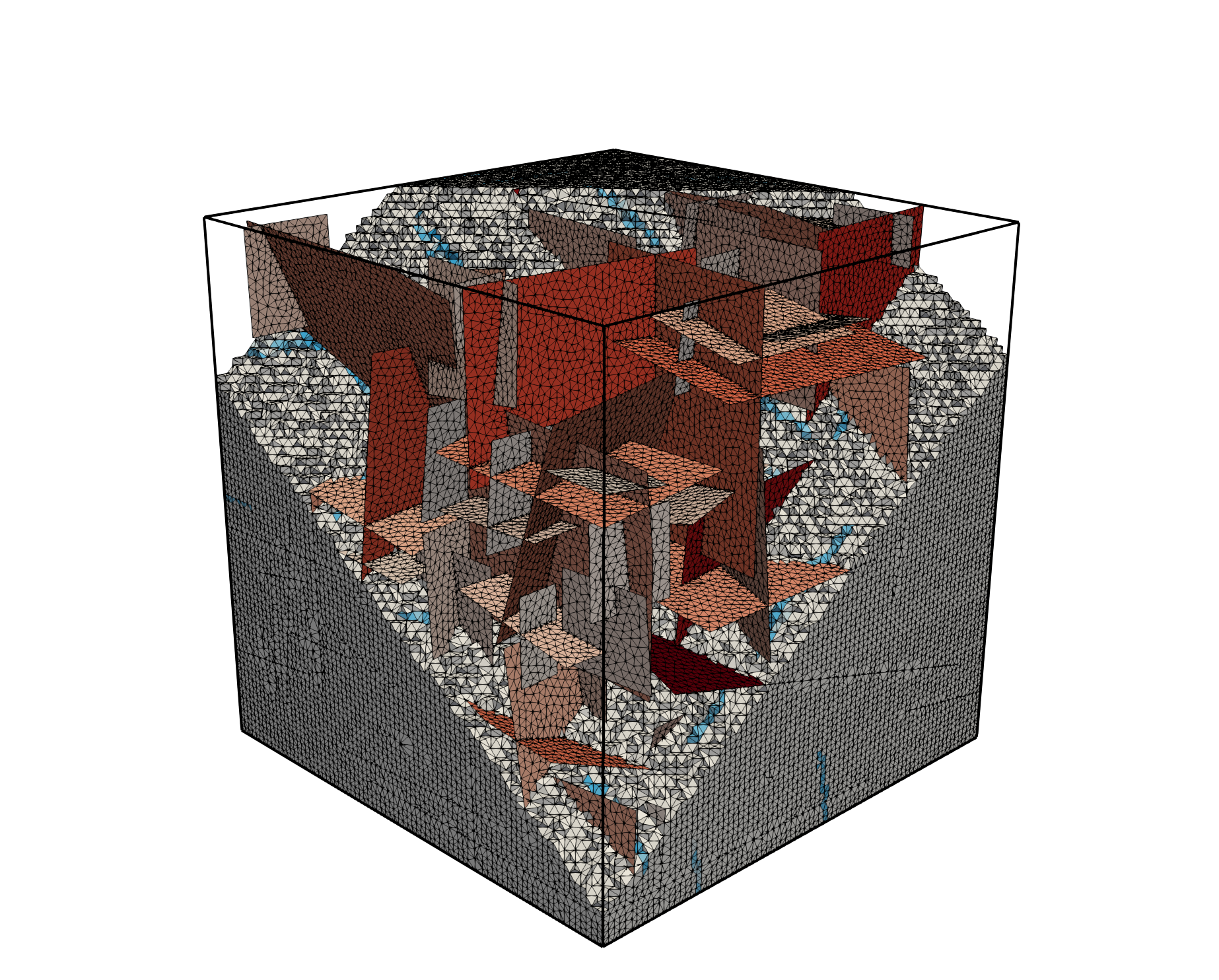} 
 \end{tabular}}
\caption{\label{fig:example_mesh}Resulting DFM mesh produced by extracting the primary subnetwork (red fractures) and upscaling the secondary structure whose affect on flow is captured via upscaling. The DFM volume mesh contains 2,902,906 tetrahedrons and 499,333 vertices. The primary subnetwork of the DFN is meshed with 53103 vertices and 103068 cells.}
\end{figure}

Figure~\ref{fig:dfn_mesh} (bottom left) shows network backbone which contains 135 fractures and the secondary structure is shown in Figure~\ref{fig:dfn_mesh} (bottom right). 
This particular backbone was identified by recursively removing all dead-end fractures, i.e., isolating the 2-core of the graph based on the DFN~\cite{hyman2017predictions}.
Notice that the backbone is much simpler than the entire network. 
The DFN backbone is meshed with 53103 vertices and 103068 cells, a reduction of about 30\%.
This method is one of the least aggressive backbone identification methods and retains a large portion of the network.
Other methods have been used to reduce the network to less than 5\% of the original number of fractures with good agreement in particular transport attributes~\cite{hyman2017predictions,hyman2018identifying}.
This network reduction/simplification facilitates the automated mesh generation of the surrounding matrix, which is shown in Fig.~\ref{fig:example_mesh}.
The resulting volume mesh contains 2,902,906 tetrahedrons and 501,298 vertices. 
The blue cells in the mesh correspond to the variable permeability of the upscaled matrix cells, which effectively capture the secondary structure of the DFN. 
While we used the method presented by Sweeney et al.~\cite{sweeney2019upscaled}, any upscaling method that works with unstructured tetrahedrons can be employed.
Note that the secondary structure does not need to be meshed, and there are no constraints on its size or complexity.
This aspect thereby further reduces the computational burden of the overall method to create the hybrid mesh.

The principal parts of the methods are described in Algorithm 1.
\begin{algorithm}\label{alg:high-lelvel}
\caption{DFM generation and meshing}
\begin{algorithmic}
\Ensure: 
\begin{enumerate}
\item Generate a two-dimensional discrete fracture network $\Omega_f \in \mathbb{R}^2$
\item $\Omega_f$ is partitioned into primary $\Omega_f^\prime$ and secondary $\Omega_f^\ast$ subnetworks such that $\Omega_f = \Omega_f^\prime \cup \Omega_f^\ast$ and they are disjoint $\Omega_f^\prime \cap \Omega_f^\ast$.
\item  The primary fracture network $\Omega_f^\prime$ is meshed with a uniform resolution Delauney triangulation that conforms to the fracture intersections denoted $\widehat{\Omega_f^\prime}$.
\item  The rock matrix $\Omega_m$ is meshed using Delauney tetrahedrons, where the faces of the tetrahedrons conform to the mesh of the fractures denoted $\widehat{\Omega_m}$. 
\item  The secondary network is upscaled into effective properties in the matrix volumes $\Omega_f^\ast \rightarrow \widehat{\Omega_m}$.
\item  $\widehat{\Omega} = \widehat{\Omega_m} \in \mathbb{R}^3 \cup \widehat{\Omega_f^\prime} \in \mathbb{R}^2$ - A multi-dimensional mesh representation of the domain is discretized using mimetic finite differences for the governing equations for flow and transport on the multidimensional system $\widehat{\Omega_m}$ and $\widehat{\Omega_f^\prime}$.
\end{enumerate}
\end{algorithmic}
\end{algorithm}

\section{Methodology Verification and Exposition}\label{sec:examples}

In this section, we provide three examples of the method. 
The numerical methods presented in this work have previously been verified for single-phase flow and transport problems in DFM systems in \cite{berre2021verification}.
In that study, four benchmark cases for single-phase flow and transport in three-dimensional fractured porous media were given to 11 participating groups, including the Amanzi team. 
In each of the problems, Amanzi successfully produced accurate results for the solutions. 
However, that analysis was completed using provided mesh files and not the meshing algorithm developed in this work.
As a result, we first present verification for transport in a single fracture coupled to a matrix here at various resolutions for further completeness. 
Next, we consider a smaller network composed of five fractures. 
Thereon, we resolve flow and transport under a set of different fracture and matrix properties ratios to demonstrate the robust nature of the method. 
In the final example, we use the larger network provided in section~\ref{sec:backbone_example} that originally contains 257 fractures before the backbone is calculated, which contains 135 fractures. 
Here, we demonstrate the scalability of the method to more complicated networks and the effectiveness of the backbone/upscaling method.

\subsection{Example 1: Single Fracture}

\begin{figure}[htb!] \centerline{
 \begin{tabular}{cc}
\includegraphics[width=0.95\textwidth]{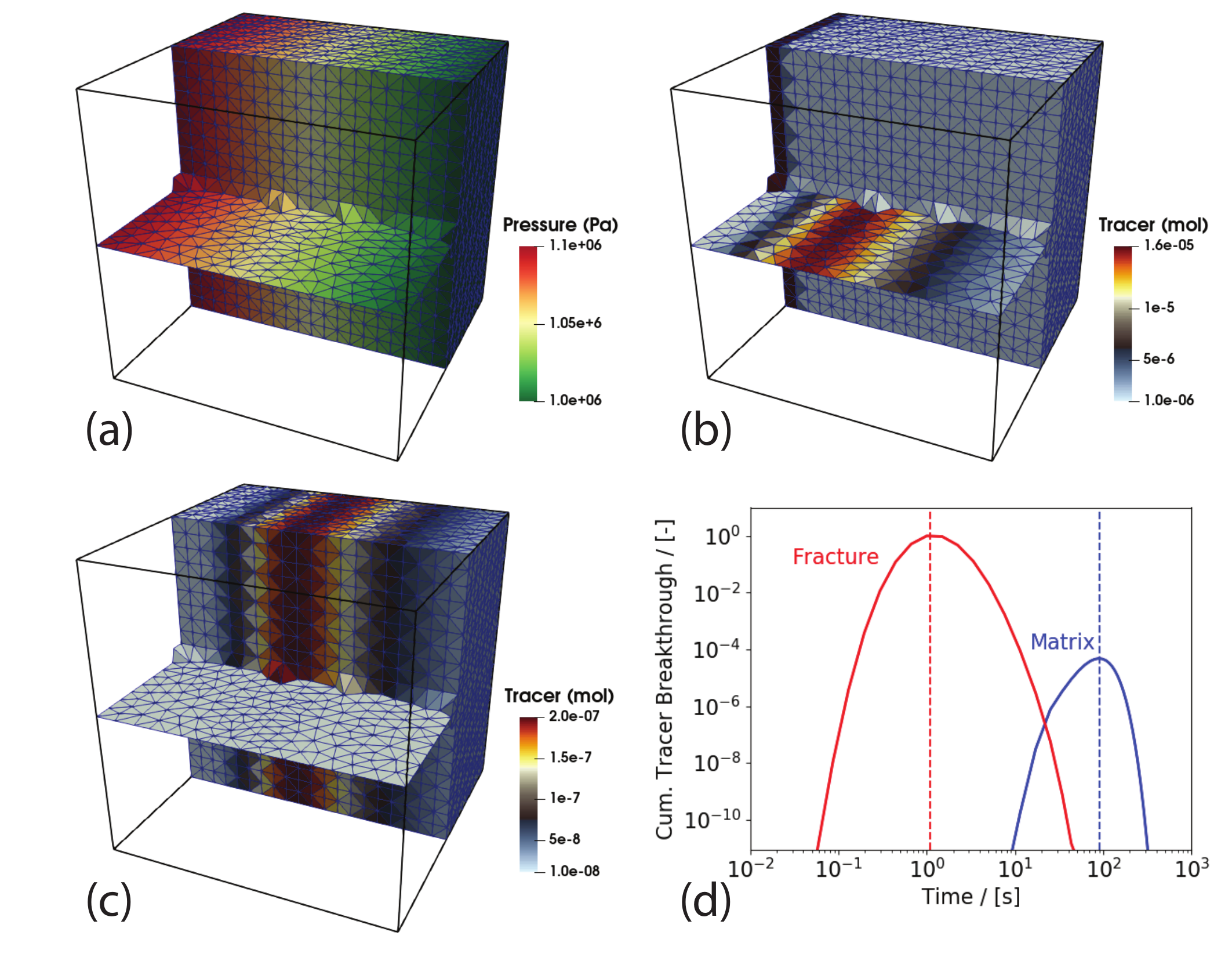} 
 \end{tabular}}
\caption{\label{fig:single_frac} Verification simulation of a single fracture coupled to matrix. (a) Steady state flow solution. (b) Early time transport of solute in fracture. (c) Late time transport of solute in matrix. (d) Breakthrough curves of solute in both fracture domain (red) and matrix domain (blue) \jeffrey{FOR MATT: Could label the figure directly and label the red curve "Fracture" and the blue curve "Matrix"}. Analytical expected values calculated from Darcy's Law are shown by vertical lines.}
\end{figure}

In this particular example, shown in Figure~\ref{fig:single_frac}, the single fracture is parallel to the direction of flow. We prescribed Dirichlet pressure boundary conditions of $1.1 \cdot 10^6$ Pa to the inlet face and $10^6$ Pa to the outlet face. 
In both cases, the fracture and matrix domain are given the same boundary conditions. The porosity of both the fracture and the matrix are arbitrarily chosen as 1.0 and 0.01, respectively. 
The permeability of the matrix is isotropic and given a value of $10^{-15}$ m$^2$. 
The fracture aperture is $10^{-5}$ m, hence the isotropic permeability is given by the cubic law, and 
is $8.33 \cdot 10^{-12}$ m$^2$. The domain is 1 m$^3$.

The transport of passive solute in the domain is modeled using the ADE equation~\eqref{eq:multi-transport-discrete} once a steady state pressure solution is achieved, which is shown in Figure~\ref{fig:single_frac}a. 
We simulate a pulse of solute injected into both the fracture and matrix domains at $t=0$ s using a Heaviside function for $0.5$ s. 
The expected arrival time of the solute at the opposite boundary to the inlet face can be predicted using Darcy's Law. 
In this case, we would expect two peaks of solute in the breakthrough PDFs, one at $t = 1200$ s due to the solute breakthrough from the fracture, and the other at $t = 10^5$ s due to the solute breakthrough from the matrix.

Figure~\ref{fig:single_frac} shows the flow solution (a), two snapshots of solute transport (b) and (c), and normalized breakthrough curves of the solute (d). 
In (d), the red line corresponds to transport in the fracture and the blue line corresponds to transport in the matrix. The curves are normalized to the peak breakthrough in the fracture. 
It is clear that the peaks agree with the analytical expected values (the vertical lines). 
In each case there is numerical diffusion due to the discretization and order of the transport scheme. 

\subsection{Example 2: Five fracture Network}

\begin{figure}[htb!] 
    \centerline{
        \includegraphics[width=0.95\textwidth]{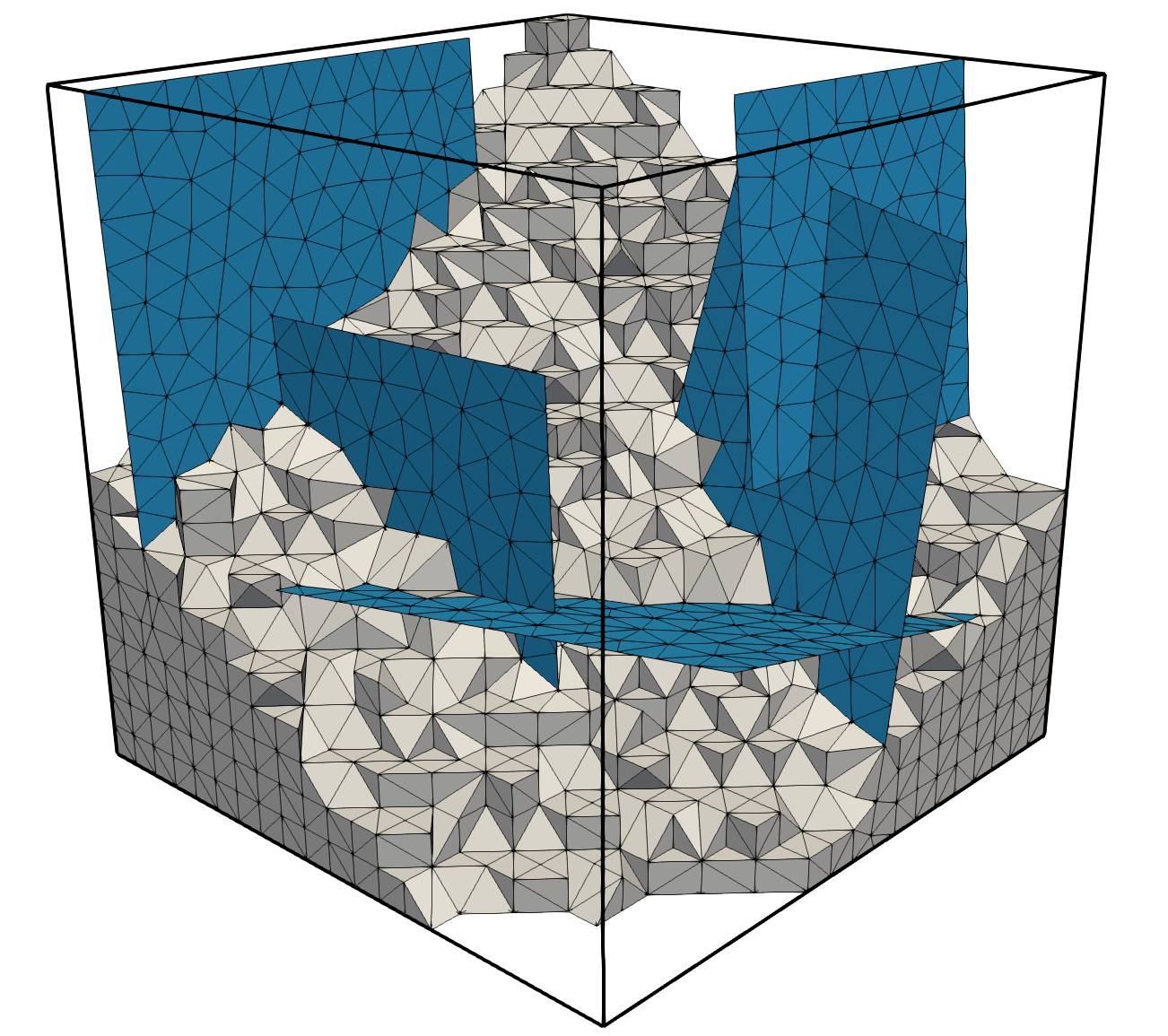} 
    }
\caption{\label{fig:5_fracture} Example network composed of five fractures. There are no regions of interference in this network, and the matrix mesh is perfectly conforming to the DFN mesh even though the fractures are not aligned with the background hexahedron mesh. }
\end{figure}

Our next example network is made up of five square fractures within a 10-meter cube, shown in Figure~\ref{fig:5_fracture}.
Each fracture has a side of length 4 meters shown in blue.
The fracture orientations are not orthogonal to one another and their intersections form obtuse and acute angles. 
Therefore, they are not aligned with the background tetrahedral mesh.
The mesh has a uniform resolution with triangle edge length of 0.5 meters, which corresponds to an $h$ value of 1 m. 
The mesh of the DFN has 1099 cells and 628 vertices and is colored black. 
The matrix mesh is made up of tetrahedrons with an edge length of 0.5 meters that is made up of 17861 cells and 3847 vertices.
There are no regions of interference in this network, and the matrix mesh is perfectly conforming to the DFN mesh. 

\subsubsection{Transport Verification}

\begin{figure*}[htb!] \centerline{
 \begin{tabular}{cc}
\includegraphics[width=0.65\textwidth]{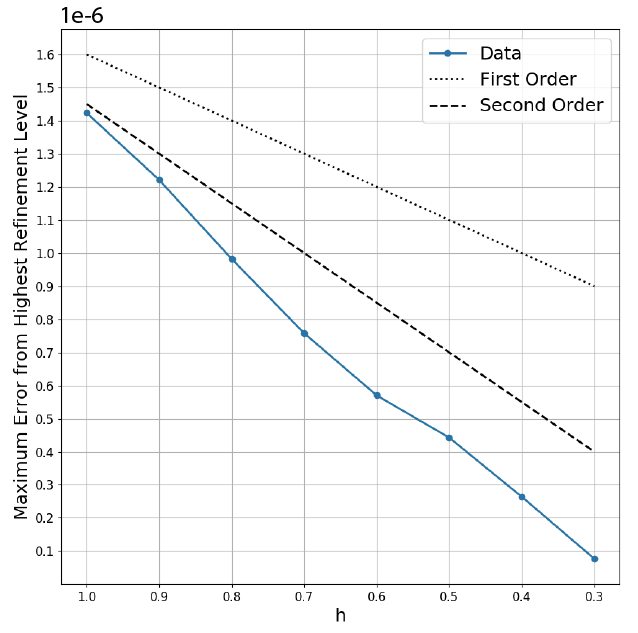}
 \end{tabular}}
\caption{Transport convergence/second-order verification in the network shown in Fig.\ref{fig:5_fracture}. Additional mesh information is presented in Table~\ref{tbl:meshes}. \label{fig:5_frac_converg} }
\end{figure*}

Note that the mesh here is not grid aligned as is the previous verification test.  
Specifics of the meshes are reported in table~\ref{tbl:meshes}.
To verify the second order-accuracy of the transport scheme, we compare the tracer breakthrough curves at the outlet of the domain. 
We set the fracture permeability to be greater than the matrix permeability: $k_f = 8.3\cdot 10^{-10}$ m$^2$, which corresponds to an aperture of $10^{-4}$ m, and matrix permeability of $k_m = 10^{-15}$ m$^2$. 
We consider steady flow via a pressure difference across the x-direction of the domain (0.1 MPa) and a pulse injection into the inflow face. 
For each refinement level, we take the maximum difference between that level and the finest level of refinement (h = 0.25 m) over the entire transport breakthrough curve at the outlet face.
The results are plotted in Fig.~\ref{fig:5_frac_converg} along with lines showing first order and second order convergence. 
The data matches the second order convergence line thereby confirming the derivation and implementation of the numerical scheme for transport.

\begin{table}
\label{tbl:meshes} 
\centering
\caption{Information about various mesh resolutions for the 5 fracture example shown in Fig.\ref{fig:5_fracture}. Convergence plots for transport though the DFM are shown in Fig.\ref{fig:5_frac_converg}}
\begin{tabular}
{llll}
\hline
 h &  \# DFN Triangles & \# DFM Tetrahedrons  &  Maximum Error from h = 0.25m \\ \hline
 1  &  1088 & 44189 & 1.18e-6 \\
0.9  & 1364 & 59111 & 1.04e-6 \\
0.8  & 1695 & 88136 &  8.56e-7  \\
0.7  & 2183 & 124345  &  6.80e-7\\
0.6  & 2970 & 202275  &  5.35e-7 \\
0.5  & 4266 & 369361 &  4.12e-7\\
0.4  & 6674 & 716722 &  2.69e-7 \\
0.3  & 11738 & 1686799 &  8.99e-8 \\
0.25  & 16965 & 2991156 & -  \\
\end{tabular}
\end{table}

\subsubsection{Variable Hydraulic Properties}

\begin{figure*}[htb!] \centerline{
 \begin{tabular}{cc}
\includegraphics[width=0.95\textwidth]{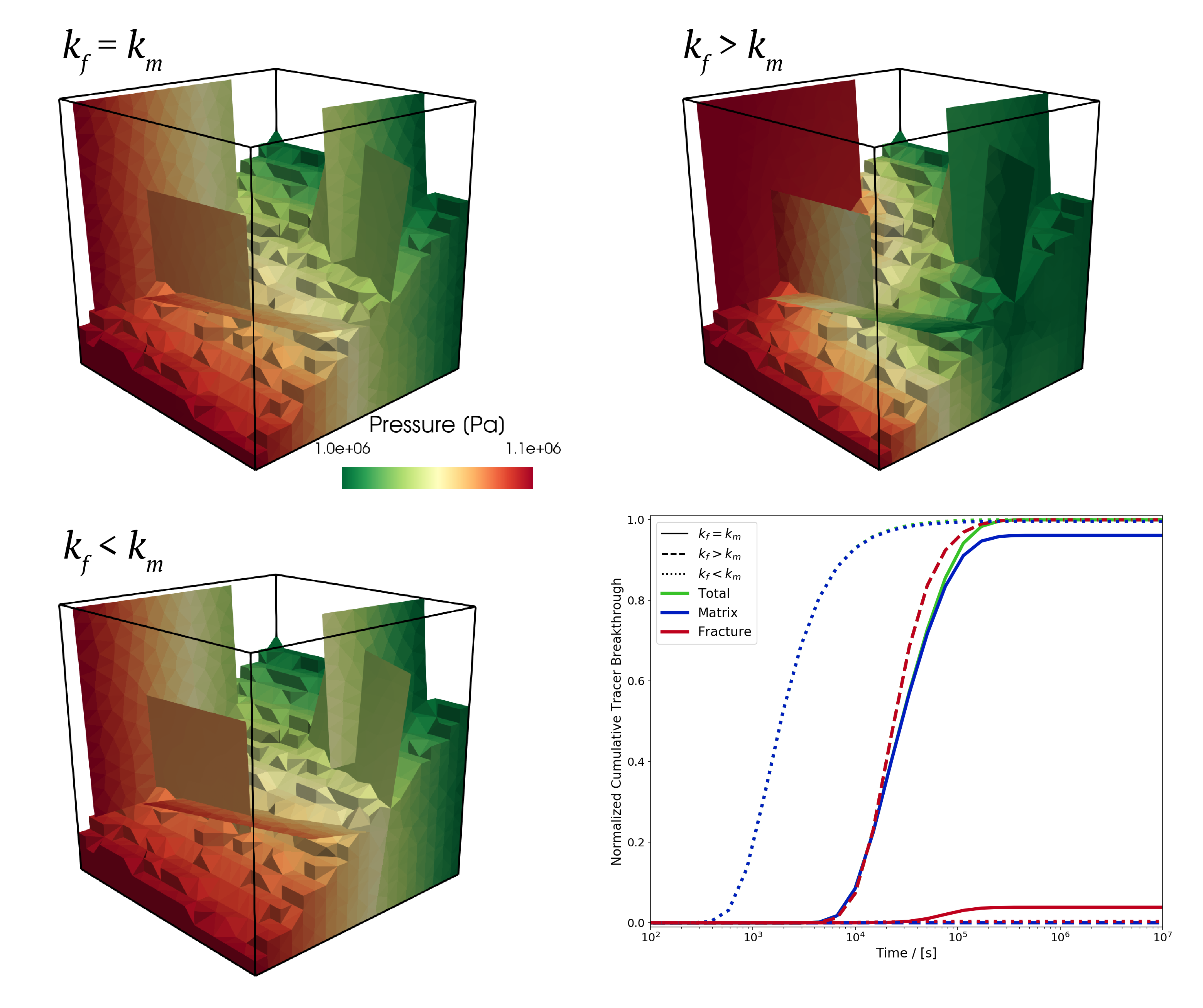}
 \end{tabular}}
\caption{ Network made up of five square fractures embedded within a permeable matrix. (Upper Left) Fractures and matrix has the same permeability. (upper right) Fractures are more permeable than the matrix. (lower left) Fractures are less permeable.  (lower right) Solute Transport breakthrough curves in the three scenarios.  \label{fig:DFM_small} }
\end{figure*}

We perform three flow and transport simulations with various contrasts between fracture and aperture permeability. 
We perform these simulations on the h = 1 m case. 
In all cases, there is steady flow created by a pressure difference across the domain aligned with the x-axis. 
The fluid is set to be isothermal water.
In all cases, we fix the fracture porosity at 1.0 and the matrix porosity at 0.1. 
In the first case, we set the fracture permeability to be equal to the matrix permeability: $k_f = k_m = 8.3\cdot 10^{-12}$ m$^2$, which corresponds to an aperture of $10^{-5}$ m. 
In the second case, we set the fracture permeability to be greater than the matrix permeability: $k_f = 8.3\cdot 10^{-10}$ m$^2$, which corresponds to an aperture of $10^{-4}$ m, and matrix permeability of $k_m = 10^{-15}$ m$^2$. 
In the third case, we set the fracture permeability to be less than the matrix permeability: $k_f = 10^{-15}$ m$^2$, which corresponds to an aperture of $10^{-7}$ m, and matrix permeability of $k_m = 10^{-12}$ m$^2$. 
Pressure solutions in the three scenarios are shown in Fig.~\ref{fig:DFM_small} (upper left) case 1: $k_f = k_m$, (upper right) case 2: $k_f > k_m$, and  (lower left) case 3: $k_f < k_m$.
In case 1, there is a smooth pressure gradient through the entire domain, and no irregularities caused by the presence of the fractures are observed.
In case 2, the higher permeability of the fractures results in a pressure conduit that forms through the network. 
The pressure in the matrix is altered accordingly to push flow into the fractures rather than continue through the lower permeable matrix. 
In case 3, the fractures act as a barrier. Specifically, observe the lower right corner of the domain where there is a pressure back up against the fractures and abrupt changes in the pressure through the fracture plane. 

We also consider passive solute tracer transport of a plus injection along the inflow boundary.
The normalized cumulative tracer breakthrough curves at the outlet plane for all three cases are shown in the lower right subfigure of Fig.~\ref{fig:DFM_small}. 
The line style indicates case number (case 1- solid, case 2-dashed, case 3-dotted), and the color corresponds to the region of the domain (whole face, matrix, or fractures) the tracer exits through. 
The curves are normalized so that the whole face values are equal to 100\% of the tracer at the end of the simulation. 
For case 1, the tracer exits through both the fracture and matrix. 
The fracture tracer is slightly delayed because of the higher porosity, which results in lower transport velocity. 
For case 2, all of the tracer exits through the fracture, which is to be expected with the pressure solution and simulation design. 
For case 3, all of the tracer exits through the matrix, which is to be expected given the lower permeability of the fracture.

\subsection{Example 3}

Our final example is the case of the mesh previously shown in Section~\ref{sec:backbone_example} that originally contains 257 fractures before the backbone is calculated, which contains 135 fractures. 
We consider two cases: one in which the 122 fractures not included in the backbone are upscaled to effective properties in the matrix mesh, and another in which they are not.
In each case, the multi-dimensional DFM mesh is geometrically identical and we simulate the same single phase flow and transport simulation with the same boundary and initial conditions apart from the upscaled properties. 
Specifically, we consider isothermal flow and transport of water in the $x$-direction driven by a pressure gradient where the left boundary is given a Dirichlet value of 1.1 MPa and the right boundary is given a Dirichlet value of 1.0 MPa. As in Section 4.2, transport is simulated by a pulse of non-reactive tracer.

\begin{figure*}[htb!] \centerline{
 \begin{tabular}{cc}
\includegraphics[width=0.95\textwidth]{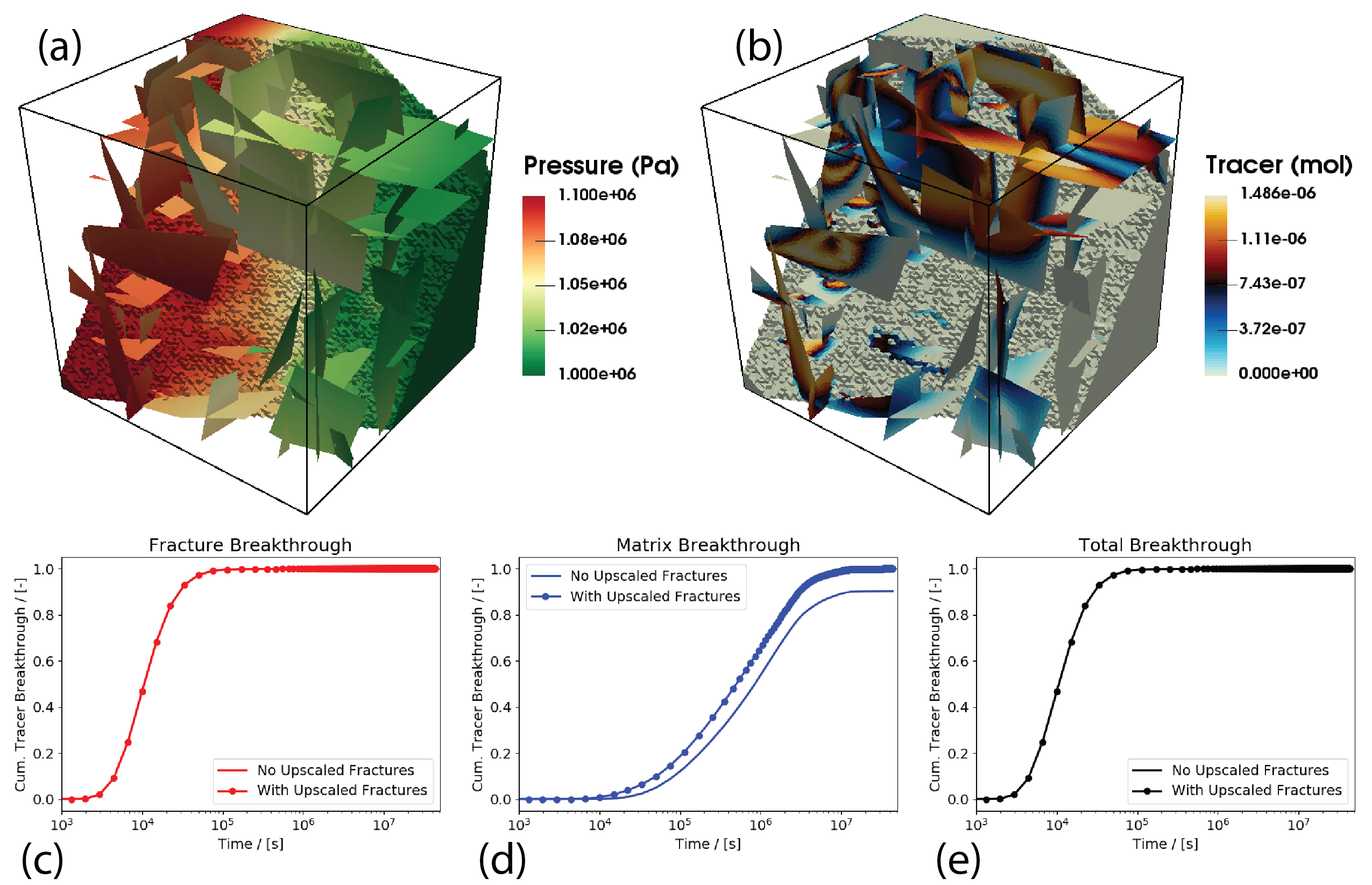}
 \end{tabular}}
\caption{ \jeffrey{FOR MATT: Might label the c,d,e plots with titles above the curves, "Fracture Breakthrough", "Matrix Breakthrough", "Total Breakthrough". remove minor ticks from legends. Fix capitalization in lengths for bottom row.} (a) Steady state flow solution (b) transport of solute at $~t=6300$ s (c) breakthrough curve of solute exiting the fracture networks (d) breakthrough curve of solute exiting matrix (e) breakthrough curve of total solute exiting domain. \label{fig:comparison} }
\end{figure*}

Fig.~\ref{fig:comparison} a,b shows snapshots of the the steady-state flow solution and transport at $~t=6300$ s for the case where upscaled fractures are included. In Fig.~\ref{fig:comparison} c,d,e the breakthrough curves of the solute through the right boundary face are plotted, comparing the two cases. 
It is clear that in each case, the breakthrough from the fracture network is equivalent, which is expected because the fracture networks are identical between the two simulations. However, there is a slight deviation in the matrix. 
When upscaled fractures are included in the matrix properties, there is a slightly earlier breakthrough of solute, as well as slightly more ($~$10\%) solute that passes through the outlet. This phenomenon can be attributed to disparate zones of higher permeability from the upscaled fractures. 
However, these differences in the matrix ultimately do not affect the total breakthrough since in this particular network the fracture network still dominates the flow and transport. 
How important it is to include the upscaled fractures will depend on the network and the hydraulic properties of the system and how aggressive of a backbone algorithm is chosen to include in the DFM.

\section{Remarks}\label{sec:remarks}

We have proposed a complete workflow to simulate single-phase flow and transport in fractured porous media using a discrete fracture matrix approach. 
The method is based on a method for mesh generation of and around a complicated 3D fracture network.
We addressed the issues of conforming mesh generation to produce a conforming Delaunay tetrahedralization of the volume surrounding the fracture network where the triangular cells of the fracture mesh are faces in the volume mesh. 
We provide two extensions of the base method to address pathological cases that commonly arise and degrade mesh quality.
The first is an extension of the feature rejection algorithm for meshing (FRAM) presented in Hyman et al.~\cite{hyman2014conforming} which ensures that all three-dimensional length scales are larger than a user-defined length-scale. 
The second method partitions the network into a primary structure, through which the majority of flow passes through and to which the mesh conforms, and the secondary structure, whose effects on flow and transport are upscaled into the spatially variable permeability field of the surrounding rock matrix.  

This high quality conforming mesh facilitates the efficient discretization of the governing equations for flow and transport, which is carried out using second-order accurate mimetic finite differencing.
The implementation of these numerical methods for high-performance computing environments is performed using a hierarchically designed multi-process coordinator and process kernel structure that allows for both strong and weak coupling of the fracture/matrix domains.
Additional physical models beyond simulate single-phase flow and transport  supported by the adopted open-source simulator include energy transport, rock and fluid compressibility, and chemical reactions. 

We provided verification tests of the method compared with analytic solutions for flow and transport. 
We also provided two expositions of the method in complex fracture networks. 
In the first, we demonstrated that the method is robust by considering scenarios where a network of five fractures is a barrier to flow,  a primary pathway, or offers the same resistance as the surrounding matrix. 
In the second example, flow and transport through a fully three-dimensional stochastically generated network containing 257 fractures are presented.
The methods presented here will be made freely available in the open-source code {\sc dfnWorks} in a future release.

\section*{Acknowledgments} 
J.D.H. and M.R.S. gratefully acknowledges support from the LANL LDRD program office Grant Number \#20180621ECR.
J.D.H, M.R.S., and J.D.M gratefully acknowledges support from the LANL LDRD program office Grant Number \#20220019DR. 
M.R.S. would also like to thank support from the Center for Nonlinear Studies.
J.D.H and M.R.S. thank the Department of Energy (DOE) Basic Energy Sciences program (LANLE3W1) for support. 
Los Alamos National Laboratory is operated by Triad National Security, LLC, for the National Nuclear Security Administration of U.S. Department of Energy (Contract No. 89233218CNA000001).
LAUR \# LA-UR-21-31458. 

\appendix

\section{Meshing Algorithm}\label{app:meshing}

\begin{algorithm}\label{alg:dfm_ps}
\caption{Algorithm to generate a conforming Delaunay tetrahedralization of a three-dimensional fracture network.}
\begin{algorithmic}
\Require $\Omega_f$ \Comment{DFN with intersection conforming Delaunay triangulation with cell edge length $\approx h$}
\Ensure $\Omega$ \Comment{DFM Delaunay tetrahedralization with cell edge length $\approx h$ and triangle faces conforming to $\Omega_f$} 
\State
\State {Step 1: Create Background Mesh}
\State{$\Omega_{m}$} \Comment{Generate a directionally isotropic tetrahedrons mesh with cell edge length $\approx h$}

\State
\State {Step 2: Excavate Background Mesh}
\ForAll {$ e \in \Omega_f$}
\State{$s(e) \in \mathbb{R}^3$} 
\EndFor  
\Comment{ For each triangle $e$ in the DFN mesh, compute the circumscribed sphere $s(e)$}

\
\If {$v$ $\in \Omega_{m}$ $\in s(e)$}
\State {$\Omega_{m^{-}} \gets$ $v \setminus \Omega_{m}$ and $E(v) \setminus \Omega_{m}$} 
\EndIf
\Comment{If a vertex $v$ of the tetrahedron mesh falls on or within any circumscribed sphere, then it is removed along with any tetrahedron associated with the vertex $E(v)$ to produce the excavated mesh $\Omega_{m^{-}}$}

\ForAll {$v \in \Omega_{f}$}
\State{$d(v,u)~\forall~u~\in~\Omega_{m^{-}}$} 
\EndFor 
\Comment {Compute the distance field between vertices of $\Omega_{f}$ and $\Omega_{m^{-}}$}
\State

\State
\State {Step 3: Create Conforming DFM Mesh}
\If {$d(v,u) \leq$ $h/2$}
\State { $\Omega_{m^{-}} \gets~u \setminus \Omega_{m^{-}}$}
\EndIf
\Comment {If any vertex in the excavated mesh is too close to the DFN, then it is removed.}

\State {$\mathcal{P}(\Omega) = \mathcal{P}(\Omega_f) \cup \mathcal{P}(\Omega_{m^{-}})$ } 
\Comment {Create point cloud that is the union of the DFN vertices and remaining points in matrix mesh}
\State

\State {$\Omega \gets \mathcal{P}(\Omega)$} 
\Comment{Reconnect the point cloud using a Delaunay algorithm to create the final mesh.}

\end{algorithmic}
\end{algorithm}

\newpage 

\bibliography{dfm}

\end{document}